\documentclass[journal,10pt]{IEEEtran}
\usepackage{tabularx}
\usepackage[noadjust]{cite}
\usepackage{tabularx}
\usepackage{diagbox}
\usepackage{amssymb}
\usepackage{amsmath}
\usepackage{multirow}
\usepackage{booktabs}
\usepackage{hhline}
{}
{}
{}
\usepackage{setspace}
\usepackage{amsthm,amsmath,amssymb}
\usepackage{mathrsfs}
\usepackage{stfloats}

\usepackage{enumerate}

\usepackage[ruled,vlined,linesnumbered]{algorithm2e}  

\usepackage{bm}   

\newcommand{\RNum}[1]{\uppercase\expandafter{\romannumeral #1\relax}}

\usepackage{booktabs}
\usepackage{xcolor}
\usepackage{balance}
\definecolor{flexred}{rgb}{0, 0, 0}

\ifCLASSINFOpdf
  \usepackage[pdftex]{graphicx}
  \graphicspath{{./figure/}}
  \DeclareGraphicsExtensions{.pdf,.jpeg,.png}
\else
\fi

\ifCLASSOPTIONcompsoc
\usepackage[caption=false,font=normalsize,labelfont=sf,textfont=sf]{subfig}
\else
\usepackage[caption=false,font=footnotesize]{subfig}
\fi

\usepackage{mathtools,amssymb}

\usepackage{cuted}
\definecolor{flexred}{rgb}{0.9, 0, 0}

\begin{document}

\title{Joint Optimization of Relay Selection and Transmission Scheduling for UAV-Aided mmWave Vehicular Networks}
\author{
Jing~Li,~
Yong~Niu,~\IEEEmembership{Member,~IEEE,}
Hao~Wu,~\IEEEmembership{Member,~IEEE,}
Bo~Ai,~\IEEEmembership{{\color{black}Fellow},~IEEE,}
Ruisi~He,~\IEEEmembership{Senior Member,~IEEE,}
Ning~Wang,~\IEEEmembership{Member,~IEEE,}
Sheng~Chen,~\IEEEmembership{Fellow,~IEEE}

\vspace*{-5mm}
}

\maketitle

\begin{abstract}
To deal with the explosive growth of mobile traffic, millimeter-wave (mmWave) communications with abundant bandwidth resources have been applied to vehicular networks. As mmWave signal is sensitive to blockage, we introduce the unmanned aerial vehicle (UAV)-aided two-way relaying system for vehicular connection enhancement and coverage expansion. How to improve transmission efficiency and to reduce latency time in such a dynamic scenario is a challenging problem. In this paper, we formulate the joint optimization problem of relay selection and transmission scheduling, aiming to reduce transmission time while meeting the throughput requirements. To solve this problem, two schemes are proposed. {\color{black} The first one is the random relay selection with concurrent scheduling (RCS), a low-complexity algorithm implemented in two steps.} The second one is the joint relay selection with dynamic scheduling (JRDS), which fully avoids relay contentions and exploits potential concurrent ability, to obtain further performance enhancement over RCS. Through extensive simulations under different environments with various flow numbers and vehicle speeds, we demonstrate that both RCS and JRDS schemes outperform the existing schemes significantly in terms of transmission time and network throughput. We also analyze the impact of threshold selection on achievable performance.
\end{abstract}

\begin{IEEEkeywords}
Millimeter-wave, vehicular networks, relay selection, concurrent scheduling, dynamic scheduling, joint optimization.
\end{IEEEkeywords}

\IEEEpeerreviewmaketitle

\section{Introduction}\label{S1}

{\color{black} Ever-growing throughput demands drive operators to exploit millimeter wave (mmWave) bands in vehicular networks, in order to improve traffic efficiency and support data-intensive services \cite{JSAC2019}. The usage of mmWave spectra is considered as a cornerstone for the next-generation communication system.
However, suffering from high path loss (PL) and frequent blockage, mmWave vehicle to vehicle (V2V) communications encounter degraded link quality and transmission interruptions \cite{mmWave1,mmWave2}. An effective way to improve the coverage, throughput and link reliability is to introduce the relay-aided transmission, with both flying unmanned aerial vehicles (UAVs) and terrestrial vehicles serving as relay nodes \cite{TCOM1}. In particular, multiple UAVs assist the ground vehicular network through air-to-ground (A2G) communications, which can offer the following three advantages over terrestrial vehicle relay.
(i)~\emph{Flexible deployment}: Following the traffic demands, UAVs can be deployed at specific sections such as crowded intersections and accident sites, mitigating terrestrial communication congestion; 
(ii)~\emph{Efficient transmission}: With controllable mobility and high altitude, UAVs have a high possibility of establishing line-of-sight (LoS) communication links with ground vehicles \cite{XMS}; 
(iii)~\emph{Connectivity enhancement}: Due to complex terrestrial propagation environment, ground-relayed vehicular links may fail, but UAVs can provide auxiliary wireless connection \cite{VTM}.}

{\color{black} 
While the integration of UAV communications brings many benefits, it is challenging to improve the performance of an aerial-ground mmWave mobile network for the following reasons. First, since power-constrained UAVs are of limited processing capacity, it is critically important to leverage aerial resources effectively, so as to maximize communication success probability, increase throughput, and achieve seamless wireless coverage.}
Moreover, there may be severe relay contentions in task-intensive regions, which increases the waiting time for requesting vehicles, harmful to delay-sensitive services including autonomous driving, computation offloading, high definition television (HDTV), etc. Besides, as both the traffic status and transmission environment in the network are changing dynamically, effective transmission schemes are particularly important to cope with the time-varying communication network. 
Therefore, how to enable rational aerial relay selection as well as efficient scheduling based on the time-varying network status to relay the service requests must be carefully studied \cite{RW1}.

\subsection{Related Works}\label{S1.1}

{\color{black}
As a promising approach to enhancing system performance, relay-aided mmWave communications have drawn substantial attention. For example, Wu \emph{et al.} \cite{N11} exploited two-hop device-to-device (D2D) relaying to improve both coverage and spectral efficiency in mmWave cellular networks. Based on the analysis of blockage effects, Ruiz \emph{et al.} \cite{9} proposed an optimum relay positioning method in the 5G framework for coverage enhancement. Moreover, Eltayeb \cite{ICCCE} innovatively introduced a relay-aided solution to correct channel estimation errors caused by imperfect antenna arrays. 
Recently, the advantages of relay usage in complex mmWave vehicular networks have been justified \cite{Glo2019}, paving the way for the construction and standardization of mmWave relay-enabled mobile networks. 
In terms of applications,} Xiao \emph{et al.} \cite{RW2} designed the relay selection and power allocation strategy for multi-source multi-relay cooperative communications, achieving significant energy savings.
{\color{black} Motivated by the potential benefits of UAV assistance, some research deployed flexible UAV relays to explore the road condition, to provide emergency communications, and to strengthen communication links among vehicles \cite{XMS, MultiU}. Specifically, the work \cite{XMS} derived a cooperative air-ground interaction network framework with multi-hop transmission. Khabbaz \emph{et al.} \cite{MultiU} proposed a UAV mobility model, based on which overall performance evaluation was conducted, shedding the light on the utilization of UAVs as external relays in vehicular networks.} 
{\color{black} However, highly dynamic vehicular environments still pose various technical challenges to the performance optimization of UAV-assisted systems, where further explorations are necessary.}

{\color{black} Another main focus in mmWave bands is the scheduling schemes for efficient transmission.} Hadded \emph{et al.} \cite{TDMA} illustrated that the TDMA technique is suitable for vehicular applications due to its collision-free and high-reliability transmission. For this reason, they further gave the classification of TDMA-based medium access control (MAC) protocols for centralized, distributed and cluster-based network topologies. In \cite{Con}, concurrent transmissions with directional antennas were proposed for spatial multiplexing, which outperformed TDMA in mmWave networks. Qiao \emph{et al.} \cite{ShenICC} proposed the multi-hop concurrent transmission (MHCT) scheme to exploit the spatial capacity of mmWave relay systems, obtaining better performance than the single-hop scheme. Besides, intensive studies \cite{NiuTCOM,TWC2020,ChenTVT} enabled link scheduling in mmWave wireless personal area networks (WPANs) and wireless backhaul networks, which are, however, constrained by static conditions. 

%

\subsection{Motivations and Contributions}\label{S1.2}

{\color{black}
Existing works have improved the network performance to some extent, but they have not fully addressed the reliability of mmWave vehicular communications, in that:}
{\color{black}\begin{itemize}
\item Relying on terrestrial relays, V2V links among remote transceivers may fail. But for UAV relaying systems, due to limited battery life, it is hard to guarantee satisfactory network performance under heavy traffic load. To date, research has seldom considered UAV-integrated two-way relaying mechanisms in vehicular networks.
\item The relative position and blockage situation among vehicles change dynamically, so that mobility prediction is required when constructing candidate relay sets and determining the relay nodes. 
\item There is a paucity of contributions that integrate the relay selection with concurrent transmission to reduce the time consumption in UAV-aided vehicular networks, which would significantly enhance transmission efficiency. 
\end{itemize}
}

To fulfill this gap, {\color{black}we employ a new candidate relay set construction method, and then solve} the joint optimization problem of relay selection and transmission scheduling in a UAV-aided mmWave vehicular network. The full-duplex (FD) scheme is enabled to further improve the transmission efficiency. Our main contributions are summarized as follows.
\begin{itemize}
\item We formulate the joint optimization problem of relay selection and transmission scheduling in a UAV-aided mmWave vehicular network into a mixed integer nonlinear programming (MINLP) problem, to minimize the time consumption while meeting the traffic demands of all flows. FD mechanism and concurrent transmission are fully exploited in the formulated problem.
\item {\color{black} To solve this problem, we develop two heuristic schemes termed random relay selection with concurrent scheduling (RCS) and joint relay selection with dynamic scheduling (JRDS). The former exhibits low complexity in algorithm design and the latter shows superior  throughput improvement. The both schemes  effectively reduce collisions in relay selection and leverage concurrent transmission.} 
\item Extensive simulations  demonstrate that JRDS achieves near-optimal performance in terms of network throughput and it outperforms other schemes in both transmission time and network throughput, {\color{black} while RCS imposes very low computational complexity.} We also analyze the impact of threshold choice on the achievable performance. 
\end{itemize}

The remainder of this paper is organized as follows. Section~\ref{S2} introduces the system model and assumption.
In Section~\ref{S3}, we formulate the optimal joint scheduling problem of relay selection and concurrent transmission, and then in Section~\ref{S4}, {\color{black}both the construction of the candidate relay set and} proposed schemes (RCS and JRDS) are described in detail. Section~\ref{S5} presents our simulation results for the achievable performance, in terms of both time consumption and network throughput. Finally, Section~\ref{S6} concludes this paper and discusses the future research directions. 
 
\begin{figure}[!b]
\vspace*{-4mm}
\begin{center}
\includegraphics[width=\columnwidth]{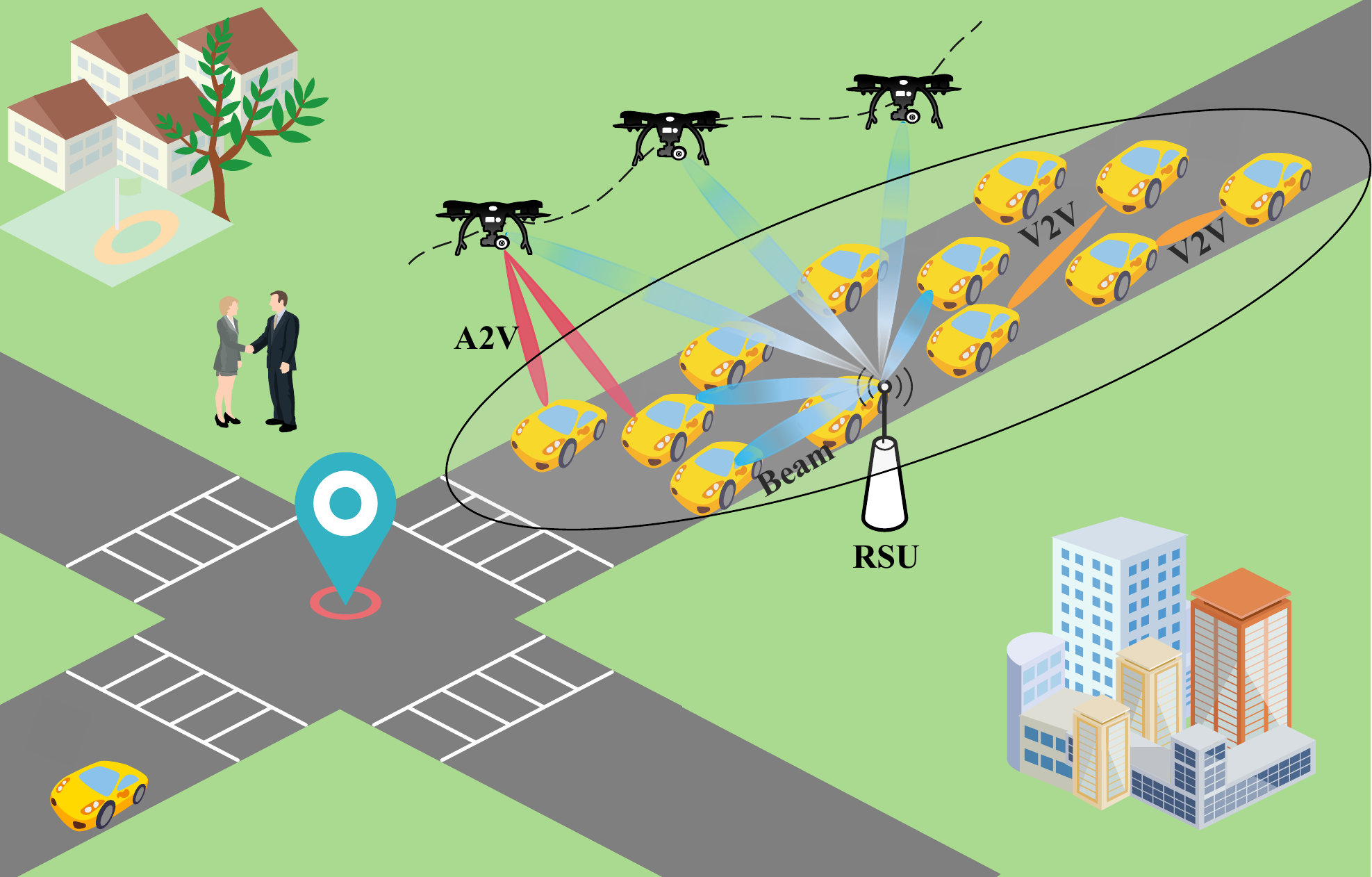}
\end{center}
\vspace*{-4mm}
\caption{A UAV-aided mmWave vehicular network.}
\label{fig:Scenario} 
\vspace*{-1mm}
\end{figure}

\section{System Overview}\label{S2}

We consider a UAV-aided mmWave vehicular network shown in Fig.~\ref{fig:Scenario}, where a roadside unit (RSU) maintains wireless connections with both vehicles and UAVs in its communication range and conducts centralized control. During mobility, there are transmission requirements of {\color{black}$N$} data flows among the vehicles, but some of them fail to be transmitted through direct V2V links due to the blockage. For a blocked flow, it can be forwarded by another vehicle or by a UAV at most once, i.e., only two-hop relay paths are exploited. Therefore, three kinds of transmission paths coexist in the network, i.e., direct links, vehicle-relayed links and UAV-relayed links. 

{\color{black} To enhance system capacity, terrestrial devices are equipped with multiple antennas and operate in full-duplex (FD) mode, which allows for simultaneous transmission and reception in the same frequency band but cannot simultaneously serve as the transmitters or receivers of multiple flows \cite{TVTK,JingLee}. But UAVs can only operate in half-duplex (HD) mode. In the transmission, we divide the time resource into $M$ equal slots and each slot has duration $T$, where concurrent scheduling can be enabled in separate time slots, to achieve time saving.} We now introduce mobility models, the antenna pattern, communication models, and the dynamic scheduling scheme. 

\subsection{Mobility Models}\label{S2.1}

Mobility modeling in mmWave vehicular networks plays a vital role in performance evaluation and designing efficient scheduling schemes. Therefore, we first provide the mobility models for both vehicles and UAVs.


\subsubsection{Vehicle Mobility Model}  

{\color{black} Assume that the arrival of vehicles follows a Poisson Process.} To avoid collisions, we determine the safe distance between the front vehicle from the following one as $\max\{2\,\text{m}, d_n\}$, where $d_n$ follows an exponential distribution with the parameter of $\overline{v}\cdot 2\,\text{s}$, and $\overline{v}$ is the average velocity [m/s] of all vehicles \cite{3GPP37.885}.

\subsubsection{UAV Mobility Model}

Assume that the coverage radius of UAV is 500\,m and several UAVs hover above the road, {\color{black} serving different regions. Each UAV moves circularly with a constant speed $V_u$ at a fixed height $h_u$, which returns to its initial location after a complete period \cite{ZR}.} Since UAVs are of limited battery capacity and processing ability \cite{HM,Lee}, the following constraints are imposed
\begin{equation}\label{equ:U1} 
P_u(t)\leq\widetilde{P_{u}}, \quad \forall{\,u,t} ,
\end{equation}
\begin{equation}\label{equ:U2} 
\frac{1}{M}\sum\limits_{t=1}^{M}P_u(t)\leq\overline{P_{u}},  \quad \forall{\,u},
\end{equation}
where $P_u(t)$ denotes the transmitting power of UAV $u$ at time slot $t$, while $\widetilde{P_{u}}$ and $\overline{P_{u}}$ are the peak and average power of UAV $u$, respectively.

\subsection{Antenna Pattern}\label{S2.2}

Directional transmission with beamforming not only compensates for the propagation loss of mmWave signals, but also provides an opportunity for high-rate communications in V2V scenarios. Therefore, we adopt directional antennas at all transceivers to accomplish sharp beamforming.


Both UAVs and vehicles are equipped with multiple antennas. {\color{black} Each vehicle (working in FD mode) supports simultaneous transmission and reception of messages at each time slot, while each UAV (working in HD mode) can only send or receive messages at a time slot.} To illustrate the antenna gain of each link, we {\color{black} give the interference model for V2V communications, with $s_i$ and $r_i$ being the source and destination of flow $i$, and $r_j$ denoting the destination of flow $j$. Specifically, during the transmission of the desired link $(s_i,r_i)$,} antennas can achieve the maximum beamforming gain {\color{black}$G_{0}$} by adjusting their directions. {\color{black} Meanwhile, $r_j$ receives the interference from link $(s_i,r_i)$.} By representing the angle deviating from the desired direction as $\theta_{r_j;s_i,r_i}$, $r_j$'s antenna gain at direction $\theta_{r_j;s_i,r_i}$ can be specified by \cite{Antenna}
\begin{equation}\label{equ:U1a} 
G_{r_j;s_i,r_i}={\color{black}G_{0}}-\min\left\{\left(\frac{\theta_{r_j;s_i,r_i}}{\theta_{\rm 3dB}}\right)^2, 26\right\} ~ \text{[dBi]},
\end{equation}
where $\theta_{\rm 3dB}$ is the half-power beamwidth. This interference model is also adopted to UAV-to-vehicle (U2V) links.

\subsection{Channel Models}\label{S2.3}

{\color{black} In this subsection, different channel models for V2V and U2V communications are considered.}

\subsubsection{V2V Links}

{\color{black} As the Nakagami-$m$ fading channel is widely used in vehicular communications and short-range communications, it is considered for V2V links \cite{Access,WYB}.} As such, the probability density function (pdf) of the channel envelope $A$ satisfies the {\color{black}Nakagami-$m$} distribution given by 
\begin{equation}\label{equ:V2VA} 
f_{A}(a;m,\varepsilon)=\frac{\varepsilon^{-m}e^{m-1}}{\Gamma(m)}\exp\left(-\frac{a}{\varepsilon}\right), m>0,\varepsilon>0,
\end{equation}
where $\Gamma(m)$ denotes the gamma function with {\color{black}the depth parameter $m$}, and $\varepsilon$ is the scale parameter controlling the spread. Hence the channel power gain $G$ obeys the Gamma distribution with parameter $m$, the pdf of which is expressed as \cite{Nakagami,Nakagami2}
 \begin{equation}\label{equ:V2VP} 
f_{G}(g;m)=\frac{\left(\frac{1}{m}\right)^{-m}g^{m-1}e^{-mg}}{\Gamma(m)}, g>0, m>0.
\end{equation}

{\color{black} For the direct link ($s_{i}, r_{i}$),} according to the LoS PL model for mmWave signals \cite{3GPP}, the received signal power at $r_{i}$ is given by
\begin{equation}\label{equ:pl1} 
P_r(s_{i},r_{i}) = k_v P_t G_0 g(s_{i},r_{i}) d^{-\alpha_v}_{s_{i},r_{i}} ,
\end{equation}
where $k_v\! =\! \left(\frac{\lambda}{4\pi}\right)^{\alpha_v}$ with $\lambda$ being the wavelength and $\alpha_v$ the PL exponent.
{\color{black} In (\ref{equ:pl1}), $P_t$, $G_0$, $g(s_{i},r_{i})$ and $d_{s_{i},r_{i}}$ respectively denote the transmission power, the maximum antenna gain, the channel power gain with Gamma distribution (\ref{equ:V2VP}), and the distance between $s_{i}$ and $r_{i}$.}

{\color{black} As concurrent transmission is exploited to improve network performance, the mutual interference (MUI) received at vehicle $r_{i}$ from $(s_{j},r_{j})$ is}
\begin{equation}\label{equ:pl2} 
P_r(s_{j},r_{i}) = k_v P_t G_{r_{i};s_{j},r_j}g(s_{j},r_{i}) d^{-\alpha_v}_{s_{j},r_{i}} ,
\end{equation}
where $G_{r_{i};s_{j},r_{j}}$ denotes the antenna gain at vehicle $r_{i}$ from link $(s_{j}, r_{j})$, and $g(s_{j},r_{i})$ represents the channel power gain of link $(s_{j}, r_{i})$, which obeys the Gamma distribution (\ref{equ:V2VP}). 

{\color{black} Moreover, due to the FD mode, the self-interference (SI) signal component arises by the transmitted signal to the received signal at the same vehicle when the same channel is used for receiving and transmitting. After SI suppressing, the residual SI (RSI) can be calculated by $\beta P_t$. Herein, $P_t$ denotes the transmit power of the vehicle, while $\beta$ indicates the SI cancellation level, which takes a value between -70 and -110\,dB for different SI algorithms \cite{SI70}. Since both MUI and SI impact the signal-to-interference plus noise ratio (SINR) at $r_{i}$, we have}
\begin{equation}\label{equ:V2V} 
\text{SINR}_{s_i,r_i} = \frac{P_r(s_i,r_i)}{N_0W +\!\! \sum\limits_{j \in \{1,\cdots ,N\}\setminus \{i\}}\!\! \big( {\color{black} a_i^{j} P_r(s_j,r_i)} + b_i^j \beta P_t\big)} ,
\end{equation}
{\color{black} in which $N_0$ is the noise power spectral density and $W$ is the channel bandwidth, while $a_i^j$ and $b_i^j$ are binary variables indicating whether flow $i$ receives the MUI and SI from flow $j\! \in\! \{1,\cdots, N\}\setminus \{i\}$, respectively. In (\ref{equ:V2V}), $\sum\limits_{j \in \{1,\cdots,N\} \setminus \{i\}} a_i^{j} P_r(s_j,r_i)$ and $\sum\limits_{j \in \{1,\cdots ,N\} \setminus \{i\}}{b^j_i\beta P_t}$ are the total MUI and SI received at vehicle $r_{i}$, respectively.} Then the achievable data rate of link $(s_{i}, r_{i})$ can be estimated by Shannon capacity \cite{Shannon}
\begin{equation}\label{equ:Q_V} 
R_{s_i,r_i} = \eta W\log\left(1+\text{SINR}_{s_i,r_i}\right) ,
\end{equation}
where $\eta\! \in\! (0, ~1]$ is the efficiency of the transceiver design.

\subsubsection{U2V Links}

As UAVs are deployed to forward blocked terrestrial flows at an altitude of $h_u$, LoS links are likely to be established between UAVs and vehicles \cite{RicianUAV}. In this case, U2V links are assumed to experience large-scale PL fading and small-scale Rician fading accounting for LoS and non-LoS (NLOS) components.
Hence, the downlink SINR of UAV $u$ to vehicle $r_k$ (denoted by link $k$ in (\ref{equ:A2G1})) can be expressed as
\begin{equation}\label{equ:A2G1} 
\text{SINR}_{u,r_k} = \frac{k_u P_uG_0d^{-\alpha_u}_{u,r_{k}}\Omega_{u,r_{k}}}{N_0 W +\!\! \sum\limits_{w\in\{1,\cdots , N\} \setminus \{k\}}\!\! \big( a_k^{w} P_r(s_w,r_k) + b^w_k\beta P_t\big)} ,
\end{equation}
where $k_u\! =\! \left(\frac{\lambda}{4\pi}\right)^{\alpha_u}$ with $\alpha_u$ being the link's PL exponent, $P_u$ denotes the transmit power of UAV $u$, and $d_{u,r_k}\! =\! \sqrt{(x_{r_k}-x_u)^2+(y_{r_k}-y_u)+h_u^2}$ is the distance between $u$ and $r_k$, with $(x_{r_k},y_{r_k},0)$ and $(x_{u},y_{u},h_u)$ being the 3D coordinates of $r_k$ and $u$, respectively. $\Omega_{u,r_{k}}$ describes the small-scale power fading, which follows a non-central $\chi^2$-distribution with the pdf given by \cite{TCOM2018}
\begin{equation}\label{equ:p10} 
\begin{aligned} 
f_{\Omega_{u,r_{k}}}(\omega) =&\, \frac{(\mathcal{K}+1)e^{-\mathcal{K}}}{\overline{\Omega}_{u,r_{k}}}\exp\left(\frac{-(\mathcal{K}+1)\omega}{\overline{\Omega}_{u,r_{k}}}\right)\\
& \times I_0\left(2\sqrt{\frac{\mathcal{K}(\mathcal{K}+1)\omega}{\overline{\Omega}_{u,r_{k}}}}\right), \,\omega\geq 0,
\end{aligned} 
\end{equation}
with $\mathcal{K}$ being the Rician factor, $\overline{\Omega}_{u,r_{k}}=1$ defined as the total power from both components, and $I_0(\cdot)$ representing the zero-order modified Bessel function of the first kind. 
Moreover, similar to (\ref{equ:V2V}), $\sum\limits_{w\in\{1,\cdots, N\} \setminus \{k\}}a_k^{w}P_r(s_w,r_k)$ and $\sum\limits_{w\in\{1,\cdots, N\} \setminus \{k\}}{b^w_k\beta P_t}$ respectively denote the MUI and SI received by link $(u,r_k)$.

As for the uplink, since UAVs fly at the height of $h_u$, the interference from terrestrial links can be ignored, and the SINR for the link from vehicle $s_k$ to UAV $u$ is given by
\begin{equation}\label{equ:G2A} 
\text{SINR}_{s_k,u} = \frac{k_v P_t G_0 d^{-\alpha_v}_{s_{k},u}\Omega_{s_k,u}}{N_0 W} .
\end{equation}

Similarly, the achievable data rates of U2V downlinks and uplinks can be estimated by Shannon capacity in (\ref{equ:Q_V}) with $\text{SINR}_{u,r_k}$ and $\text{SINR}_{s_k,u}$, respectively.

\subsection{Dynamic Scheduling}\label{S2.4}
 
\begin{figure}[!t]
\begin{center}
\includegraphics[width=0.95\columnwidth]{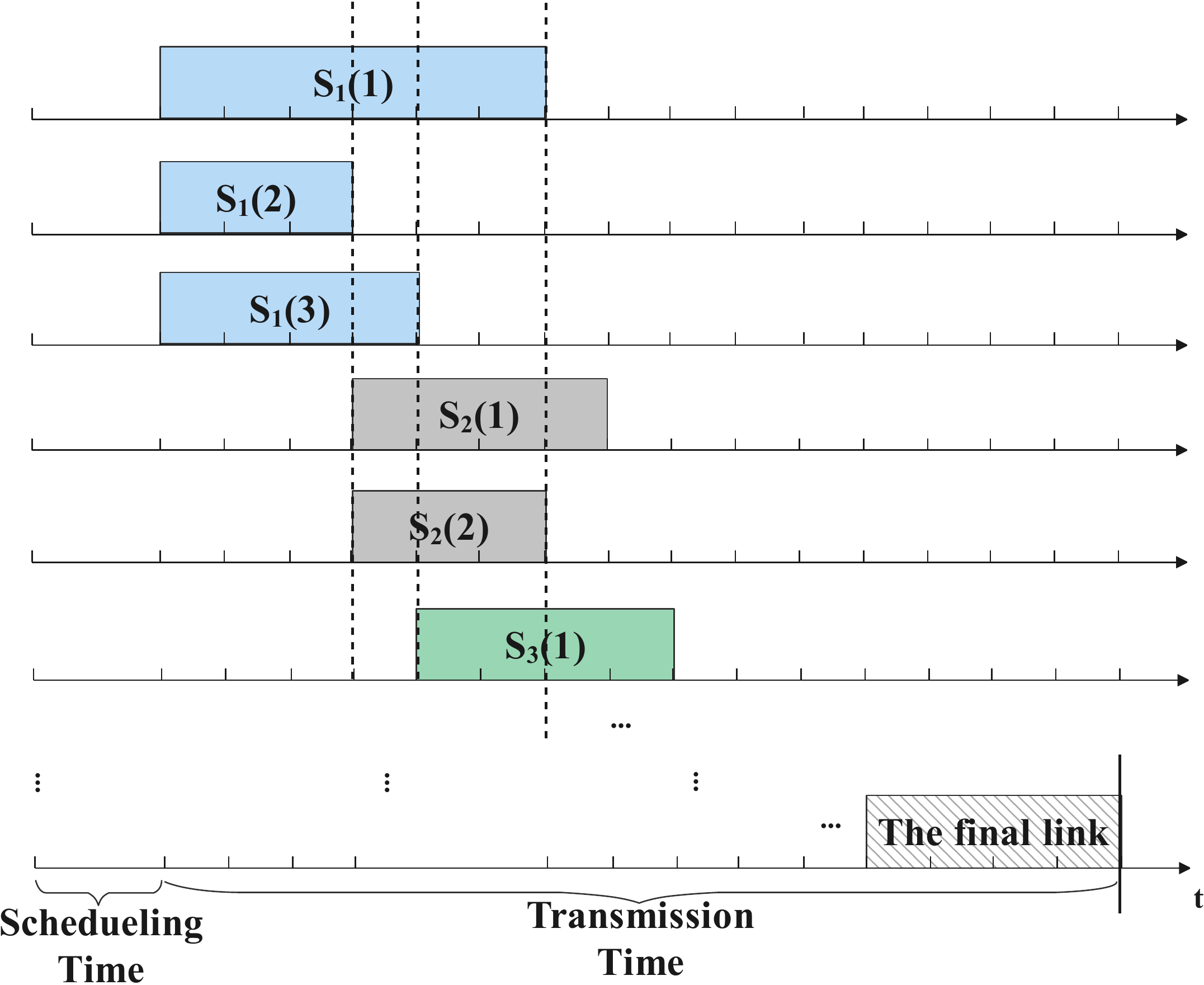}
\end{center}
\vspace*{-4mm}
\caption{Time-line illustration of dynamic scheduling.}
\label{fig:DS} 
\vspace*{-3mm}
\end{figure}

The basic idea of dynamic scheduling (DS) is illustrated in Fig.~\ref{fig:DS}. The overall time consumption is composed of a scheduling stage and a transmission stage, while the scheduling time can be amortized over multiple scheduling operations in the transmission phase \cite{DS,NiuJSAC}. The transmission time composes of multiple time slots, and multiple no-conflicting links can be transmitted concurrently in the same time interval.

During the transmission, non-adjacent direct links are scheduled to  transmit simultaneously at the beginning. Then once any of them is completed, DS is executed. Specifically, it updates the candidate relay set of each flow, regenerates contention graph and reallocates links that can be transmitted concurrently with the ongoing ones. Note that this operation repeats once there is a completed link until the final one is scheduled. As an example, $S_1$, $S_2$, $S_3$ in Fig.~\ref{fig:DS} represent the three-step dynamic scheduling. In the first step, $S_1(1)$, $S_1(2)$, $S_1(3)$ are scheduled to transmit simultaneously. When $S_1(2)$ is completed, the second step activates concurrent transmission of $S_2(1)$ and $S_2(2)$, which do not conflict with the ongoing $S_1(1)$ and $S_1(3)$. Then, when $S_1(3)$ is finished, the same operation is executed and $S_3(1)$ is activated.

\section{Problem Formulation}\label{S3}

To enhance transmission efficiency by optimal relay selection and exploiting concurrent transmission, i.e., to accomplish the traffic demand of flows with the least time intervals, we formulate a joint optimization problem for relay selection and concurrent transmission in the UAV-aided vehicular network.

Denote the candidate relay set as $\mathbf{C}$ and the scheduling as $\mathbb{S}$. We begin by  summarizing the constraints in the relay selection and transmission scheduling subproblems.

\emph{Subproblem 1:}
When flow $i$ is blocked, it can be forwarded by another vehicle or by a UAV. Therefore, how to choose its relay path is a key problem. Let $N_i$ and $N'_i$ be the binary variables that indicate whether flow $i$ is successfully transmitted by direct links and relay links, respectively. To be specific, $N_i\! =\! 1$ only when the quality of service (QoS) of a direct flow is satisfied, whereas for a blocked flow, $N'_i\! =\! 1$ only if the QoS of two hops is accommodated. Since a flow can be transmitted successfully either by a direct path or by a relay path, we have
\begin{equation}\label{equ:limit_1} 
N_i+N'_i \leq 1,\quad \forall{\,i}.
\end{equation}
We now provide the constraints on relay selection.

First, since each blocked flow can be relayed only once at most, the maximum number of hops for flow $i$ in the transmission path  satisfies $H_{\max,i}\leq 2$.

Second, the candidate relay set for flow $i$, denoted by $\mathbf{C}(i)$, may vary with time since it is affected by the relay selection of its concurrent links. {\color{black}In this case, to select the optimal relay for flow $i$, a contention function $\text{MinDegree}$ is defined to find the relay path that has the least contention with the ongoing links. Then the optimal relay for flow $i$ can be obtained as
\begin{equation}\label{equ:D} 
\mathcal R(i)=\text{MinDegree}(\mathbf{C}(i)) , \text{if}\,\, N'_i=1,  \forall{\,i}.
\end{equation}
Note that $\mathcal R(i)$ is unavailable for other blocked flows during the transmission of flow $i$.}


\emph{Subproblem 2:}
Concurrent transmissions are enabled among nonadjacent links in the scheduling. These links stay active until their traffic demands have finished. Define a binary variable $F(s^t_i, r^t_i; s^t_j, r^t_j)$ to indicate whether two links are adjacent at time slot $t$. If so, $F(s^t_i, r^t_i; s^t_j, r^t_j)\! =\! 1$; otherwise, $F(s^t_i, r^t_i; s^t_j, r^t_j)\! =\! 0$. Once one of the concurrent links is completed, we update the ${\color{black} \mathbf{C}(i)}$ for each flow, regenerate the contention graph and continue scheduling the remaining links as much as possible. We now analyze the constraints for scheduling.

First, two adjacent flows cannot be transmitted in the same time slot, i.e., {\color{black} 
\begin{align} 
& c^t_i+c^t_j\leq1,\,{\rm{if}}\, F(s^{t}_i,r^{t}_i;s^{t}_j,r^{t}_j)=1,\, \forall{\,i,j,t} , \label{equ:D-1} \\
& e^t_{ih}+e^t_{jl}\leq1,\,{\rm{if}}\, F(s^{t}_{ih},r^{t}_{ih};s^{t}_{jl},r^{t}_{jl})=1,\,\forall{\,i,h,j,l,t}, \label{equ:D-2} \\
& c^t_i+e^t_{jl}\leq1,\,{\rm{if}}\, F\big(s^{t}_i,r^{t}_i;s^{t}_{jl},r^{t}_{jl}\big)=1, \,\,\forall{\,i,j,l,t}. \label{equ:D-3}
\end{align}
}Here, $c^t_i$ is the binary variable indicating whether flow $i$ is transmitted in time slot $t$ (If so, $c^t_i\! =\! 1$; otherwise, $c^t_i\! =\! 0$), while {\color{black}$s^{t}_i$ and $r^{t}_i$} denote the source and destination of the direct path for flow $i$, respectively, at time slot $t$. Similarly, the binary variable $e^t_{ih}$ indicates whether the $h$-th hop of flow $i$ is transmitted in time slot $t$ (If so, $e^t_{ih}\! =\! 1$; otherwise, $e^t_{ih}\! =\! 0$), while {\color{black}$s^{t}_{ih}$ and $r^{t}_{ih}$} are the source and destination of the $h$-th hop path for flow $i$, respectively, at time slot $t$.

Second, for a blocked flow, there is an inherent order in the transmission of its relay path, so that different hops on the same path cannot be scheduled simultaneously, i.e.,
\begin{equation}\label{equ:Inherent} 
\sum\limits_{h=1}^{H_{\max,i}}e^t_{ih}\leq 1,\quad\forall{\,i,h,t}.
\end{equation}

Third, for flow $i$ transmitted by a relay path, its $h$th hop should be scheduled ahead of the $(h+1)$th hop, which can be formulated as
\begin{equation}\label{equ:min_power} 
\begin{aligned} 
&\sum\limits_{t=1}^{T^\star}e^t_{ih}\geq \sum\limits_{t=1}^{T^\star}e^t_{i(h+1)}, 
\, {\rm{if}} \, H_{\max,i}>1, \\
& \quad\quad\quad\quad \forall{\,h=1\sim (H_{\max,i}-1), ~ T^\star=1\sim M}.
\end{aligned}
\end{equation}
This condition represents a series of constraints with $h$ varying from 1 to $H_{\max,i}-1$ and $T^\star$ varying from 1 to $M$.

Fourth, flow $i$ can be transmitted either by a direct link or by a relay path, i.e.,
\begin{equation}\label{equ:limit_11} 
c^t_i+e^t_{ih} \leq 1, \quad \forall{\,h=1\sim H_{max,i}, t}.
\end{equation}

Therefore, the joint optimization problem P1 for relay selection and concurrent transmission is formulated as
\begin{equation}\label{equ:P} 
 \text{P1} ~\left\{ \begin{array}{cl}
\min\limits_{\textbf{C},\mathbb{S}} & \delta(\textbf{C},\mathbb{S}) , \\
{\rm s.t.} & \text{Constraints } (\ref{equ:limit_1})-(\ref{equ:limit_11}) \text{ are met}.
\end{array} \right.
\end{equation}
Here the objective $\delta$ is the total time consumption of the relay set $\textbf{C}$ and scheduling approach $\mathbb{S}$. This is a mixed integer nonlinear programming (MINLP) problem due to the discrete nature of flow scheduling with QoS constraints. Generally, such a problem is NP-hard \cite{NiuJSAC}. Considering the difficulty in solving it directly, we propose two heuristic algorithms as the practical solutions to this NP-hard problem.

\begin{figure*}[!tp]
 \vspace*{-2mm}
\begin{center}
\includegraphics[width=0.95\linewidth]{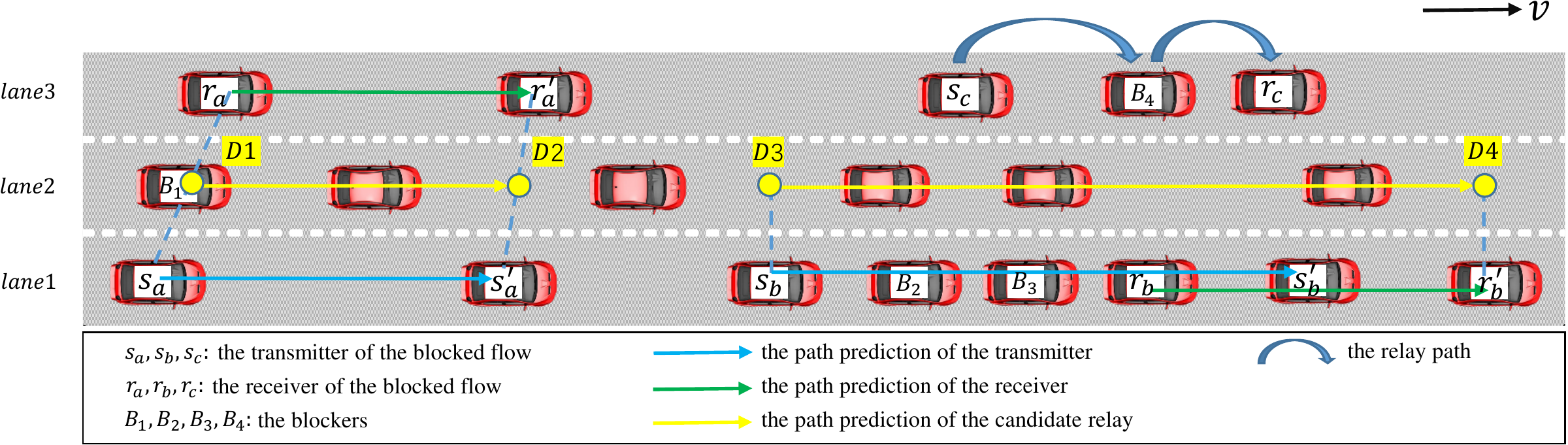}
\end{center}
\vspace*{-4mm}
\caption{Principle of relay selection.}
\label{fig:Relay} 
\vspace*{-5mm}
\end{figure*}

\section{Proposed Approach}\label{S4}

{\color{black} Since there are two key parts to be optimized in the problem P1, i.e., relay selection and concurrent transmission, we first construct candidate relay sets for blocked flows and then design two algorithms for relay selection and transmission scheduling, termed RCS and JRDS. Both direct links, vehicle-relay links and UAV-relayed links are scheduled and blocked flows are transmitted by these two algorithms.}

\subsection{Construction of Candidate Relay Set}\label{S4.1}

As shown in Fig.~\ref{fig:Relay}, we use a straight highway of three lanes as an illustration. When a flow is blocked, we need to choose an appropriate relay to forward its data. Note that the communications among vehicles that move on adjacent lanes will not be blocked, so that we conduct relay selection for the following three cases. 
\begin{enumerate}
\item For the blocked flow $a$, as its transmitter $s_a$ and receiver $r_a$ move forward on nonadjacent lanes, we select the candidate vehicle relays from the middle lane, with which flows are more likely to accomplish.
\item For the blocked flow whose transmitter and receiver move forward on the same lane, e.g., flow $b$, we select the candidate vehicle relays from its adjacent lane.
\item For the blocked flow $c$, since its transmitter and receiver move forward on the same lane and there is only one blocker between them, this blocker is used as the vehicle relay.
\end{enumerate}
The vehicle relay set obtained after this step is denoted by $\mathbf{V}'(i_{\rm flow})$, where $i_{\rm flow}$ denotes the flow concerned. To guarantee the transmission quality during relay forward \cite{Ding}, we define a parameter $\epsilon$ to obtain the candidate {\color{black}terrestrial relay set $\mathbf{V}(\bullet)$ from $\mathbf{V}'(\bullet)$}. 

More specifically, for case 1), 
\begin{equation}\label{equ:CA} 
\epsilon(a_i)=\frac{{\color{black}P(a_i)}\cap P(D1,D2)}{{\color{black}P(a_i)}},
\end{equation}
\begin{equation}\label{equ:CA-1} 
\mathbf{V(\bullet)}=\mathbf{V(\bullet )}\cup {\color{black}\{a_i\}}, {\rm{if}}\ \epsilon(a_i)\geq 0.5,
\end{equation}
where ${\color{black}P(a_i)}$ denotes the predicted moving path of each relay $a_i \in \mathbf{V}'(a)$, which can be obtained by the RSU, and $P(D1,D2)$ is the predicted moving path of the ideal relay, the length of which can be calculated by
\begin{equation}\label{equ:L} 
L(P(D1,D2))=\frac{1}{2}\left(v'(s_a)\cdot (\xi_a T) + v'(r_a)\cdot (\xi_a T)\right) .
\end{equation}
Here $v'(s_a)$ and $v'(r_a)$ are the speeds of vehicles $s_a$ and $r_a$ when flow $a$ starts transmitting, and $\xi_a$ is the number of slots that $a$ is estimated to consume, i.e.,
\begin{equation} \label{equ:CA-2} 
\xi_a =\frac{Q_a \cdot M\cdot T}{R_a \cdot T},
\end{equation}
in which $Q_a$ is the throughput of $a$, and $R_{a}$ is the rate of flow $a$ without interference from other links.

{\color{black}As for aerial UAVs, during $\xi_a$, if $P(D_1, D_2)$ is within the coverage range of some UAV, that UAV can act as a relay of $i_{\rm flow}$, by which we construct candidate aerial relay set $\mathbf{U(\bullet)}$. Therefore, we obtain the candidate relay set for $i_{\rm flow}$, where $\mathbf{C(\bullet)}=\mathbf{V(\bullet)}\cup\mathbf{U(\bullet)}$.}

Similarly, for case 2, 
\begin{equation}\label{equ:CAb} 
\epsilon(b_i)=\frac{{\color{black}P(b_i)}\cap P(D3,D4)}{{\color{black}P(b_i)}},
\end{equation}
\begin{equation}\label{equ:CA-3} 
\mathbf{V(\bullet)}=\mathbf{V(\bullet)}\cup {\color{black}\{b_i\}}, {\rm{if}}\ \epsilon(b_i)\geq 0.5,
\end{equation}
\begin{equation}\label{equ:b} 
L(P(D3,D4))=L(s_b,r_b)+v'(r_b)\cdot (\xi_b T),
\end{equation}
with $L(s_b,r_b)$ denoting the distance between $s_b$ and $r_b$. {\color{black}Also, $\mathbf{U(\bullet)}$ can be determined by judging which UAV coverage range $P(D_3, D_4)$ is in. }

For case 3, $\mathbf{V}(c)=\{B_4\}$ and $\epsilon(c)$ is a constant, {\color{black}and $\mathbf{U(\bullet)}$, $\mathbf{C(\bullet)}$ can be constructed.}


\subsection{RCS scheme}\label{S4.2}

Based on the candidate relay set, RCS provides a two-step solution to the joint optimization problem P1. First, for each blocked flow, its relay is obtained by selecting an element in its $\mathbf{C} (\bullet)$ randomly. In the scheduling phase, to avoid flow failure caused by the case that its second hop could not be scheduled in time when its first hop is completed, two hops of a flow are considered as a whole in the scheduling and the second hop can be sent immediately after completing the first one.

Next, flows are divided into groups based on the graph theory \cite{ContentionGraph}, where contention ones cannot be scheduled into the same group. To construct a contention graph $Graph(V,E)$, each flow is represented by a vertex, and there is an edge between two vertices when severe interference exists among them, i.e., relative interference (RI) exceeds the defined threshold $\sigma$ or two flows are conflicting due to the FD mode \cite{JingLee}. Note that only when neither of the two hops of a flow has an edge with another flow, there exists no edge among them. The number of edges for a vertex is defined as its degree.
 
The pseudo-code of the above flow grouping strategy is given in Algorithm \ref{alg1}. At the beginning, we input the $Group(V,E)$ for {\color{black}$N$} flows and initialize the set of grouping results $\mathbb{G} \leftarrow \emptyset$. While $V \neq \emptyset$, it calls the function \emph{generateGroup} to obtain from the flows in $V$ the ones that are not in contention with each other and form them as a new group $G\rightarrow \mathbb{G}$. Next, these flows are removed from $V$, and the edges related to these flows are also removed from $E$.  The procedure iterates until no flow left. The computational complexity of Algorithm~\ref{alg1} is on the order of ${\color{black}N^2}$, denoted as ${\color{black}\mathcal{O} (N^2)}$.

\begin{algorithm}[h]
\caption{Flow-Grouping Algorithm}
\label{alg1}  
\SetKwInOut{Initialization}{Initialization}
\KwIn{$Graph(V,E)$; \\}
\KwOut{Flow-grouped results $\mathbb{G}$; \\}
\Initialization{$\mathbb{G} \leftarrow \emptyset$;}
\While{$V \neq \emptyset$}
{	
\SetKwFunction{FGroup}{generateGroup}
  \SetKwProg{Fn}{Function}{:}{}
  \Fn{\FGroup{$V$, $E$}}{
$G \leftarrow \emptyset$\;
        \While{$V \neq \emptyset$}
{	
Obtain $v$ $\in$ $V$ with the minimum degree\;
$G=G \cup \{v\}$\;
Obtain flows $v'\in V$ that have edges with $v$\;
$V=V-v-v'$\;
	$E=E-E(v)-E(v')$\;
}    
\KwRet G\;
}
$V=V-G$\;
$E=E-E(G)$\;
$\mathbb{G} = \mathbb{G} \cup \{G\} $\;
}    
\end{algorithm}

Based on the relay selection and flow grouping results, our proposed concurrent scheduling is given in Algorithm \ref{alg:Concurrent Scheduling}, where flows in one group are transmitted concurrently while different groups are scheduled sequentially. The time slot consumption and system throughput attained by this scheme are also provided. $H_g$ is defined as 1 at the beginning, which ensures that the first hop of each flow is transmitted ahead of the second one. In lines 4-8, once one hop is completed, recalculate the data rate $R'_{g,H_g}(n)$ for the remaining flows in group $G$, in that the interference decreases due to the completion. {\color{black}Specifically, the $R'_{g,H_g}(n)$ of remaining V2V links is calculated by (9), while that of remaining U2V downlink and uplink can be estimated with SINR given by (10) and (11), respectively.} Moreover, the number of bits achieved by each flow $u_{g,H_g}(n)$ is calculated. If no hop is completed, as in lines 9-12, the data rate of each flow is not changed but $u_{g,H_g}(n)$ will increase. After that, as shown in lines 13-15, we check whether there are some hops achieving their data requirements. If the data requirement for the first hop of a relayed flow has been accomplished, increase the hop $H_g$ by 1 and repeat the iteration. As indicated in line 16, the outer loop stops when the transmission requirements for all the flows have been satisfied and $n$ is the total time slots consumed. Furthermore, we obtain the system throughput in line 17.

\begin{algorithm}[tp!]
\caption{Concurrent Transmission}
\label{alg:Concurrent Scheduling}  
\SetKwInOut{Initialization}{Initialization}
\KwIn{Flow-grouping results $\mathbb{G}$;
Throughput of each flow $Q_g$;
Number of time slots $M$;
Maximum hop of each flow $H_{\max,g}$; \\}
\KwOut{Total time slots needed after scheduling $n$;
System throughput after scheduling $U$; \\}
\Initialization{$U=0, H_g=1$;}
\For{slot $n$ ($1\leq n \leq M$)}{
 \ForEach{$G\ in \ \mathbb{G}$}{
  \While{$G \neq \emptyset$}{
   \If{one hop of some flow in $G$ is newly completed}{
    \ForEach{flow $g$ in $G$}{
     Recalculate data rate of each link $R'_{g,H_g}(n)$\;
     Calculated number of bits achieved by each flow $u_{g,H_g}(n)$\;
     go to line 13\;
		}
   }
   \Else {
     \ForEach{flow $g$ in $G$}{
      Calculated number of bits achieved by each flow $u_{g,H_g}(n)$\;
      go to line 13\;
     }
    }
    \If{any $u_{g,H_g}(n) \geq Q_g\cdot M$}{
     \If{$H_{\max,g}=2\ \& \ H_g=1$}{
      $H_g=H_g+1$\;
		}
   }
  }
 }
 break;
}
$U=Q_g\cdot {\color{black}N}/(n\cdot 0.1)$\;
\Return{$n$,\ $U$}   
\end{algorithm}

\subsection{JRDS Scheme}\label{S4.3}

Two issues that may affect the performance of the previous RCS scheme require further consideration. First, random relay selection may arise collisions among flows, which may affect the concurrent scheduling. Second, flows that are being transmitted occupy some relay nodes, which can no longer be considered as candidate relays for blocked flows at the same time. Therefore, to increase system efficiency, we design a joint solution for solving the optimization problem P1, which dynamically selects the optimal relay for each blocked flow and schedules flows in real-time. Our proposed JRDS scheme is summarized in Algorithm~\ref{alg:JRDS}.

\begin{algorithm}[tp!]
\label{alg:JRDS} 
\caption{JRDS}
\KwIn{Set of direct flows $V_d$;
Edge set $E_d$ related to $V_d$;
Set of relay flows $V_r$;
Candidate relay set to each flow $R_g$; \\}
\KwOut{Scheduling results $\mathbb{G}$;
Total time slot consumption $m$;
System throughput after scheduling $U$; \\}
  \SetKwInput{KwInitialization}{Initialization}   
  \KwInitialization{$\mathbb{G} \leftarrow \emptyset$;
	$G \leftarrow \emptyset$;\\}
  \SetKwFunction{FDGroup}{DGroup}
  \SetKwProg{Fn}{Function}{:}{}
  \Fn{\FDGroup{$V_d$, $E_d$, $G$}}{
   Obtain edges among $V_d$ and existing relay flows in group $G$, $E_{dr}$\;
  $E=E_d+E_{dr}$\;
  $DG=generateGroup(V_d,E)$\;
 \Return{$DG$\;}}
 $G=DG$\;
 \SetKwFunction{FRGroup}{RGroup}
  \SetKwProg{Fn}{Function}{:}{}
  \Fn{\FRGroup{$V_r$, $R_g$, $G$}}{
   $RG \leftarrow \emptyset$\;
   \ForEach{$v \in V_r$}{
    \ForEach{$r$ in optional relay set of flow $v$}{
     \If{$v(r)$ is not collided with flows in $(G+RG)$}
      {$RG=RG+v(r)$\;
      break\;}
   }
  }
  \Return{$RG$\;}}
 $G=DG+RG$\;
 \For{slot $m$ ($1\leq m \leq M$)}{
 {\color{black}Update the coordinates of vehicles and UAVs\;} 
 {Obtain optional relay set for each flow, $R_g$\;}
  {\If{some flow in $G$ is newly completed}
         {Obtain completed flows in $G'$\;
         $G=G-G'$\;
         $V_d=V_d-G'_d$\;
         $V_r=V_r-G'_r$\;
         $G=G+DGroup(V_d,E_d,G)$\;
         $G=G+RGroup(V_r,R_g,G)$\;
         $\mathbb{G}=\mathbb{G} \cup G$\; }
   }
   \If{flows in $(V_d+V_r)$ have been completed}
    {
      break\;
    }
  $U=Q_g\cdot {\color{black}N}/(m\cdot 0.1)$\; 
  \Return{$\mathbb{G}$,\ $m$,\ $U$}\;
 } 
\end{algorithm}
 
The algorithm starts by calling functions $DGroup$ and $RGroup$ sequentially. $DGroup$ gets the edges of unscheduled direct flows $E_d$ as well as obtains the edges among $V_d$ and the existing relay flows in group $G$, denoted as $E_{dr}$. $E_d$ and $E_{dr}$ are stored in $E$. 
Then function $generateGroup$ finds the direct flows that are not in contention with each other. Function $RGroup$ iteratively assigns suitable relays (vehicle or UAV) for each blocked flow in $V_r$, and constructs $RG$ with the ones that do not collide with the existing flows in $(G+RG)$. By now, the algorithm finds out the direct flows and relay flows that can transmit concurrently.
Then it begins the loop, which first {\color{black} update the locations of vehicles and UAVs, which affects the construction of candidate relay sets.} 
Then it detects the optional relay set for each flow at this slot, in that once completing a flow, some nodes (both vehicles and UAVs) are released and therefore become optional relays for other ones. Next, it checks whether there are completed flows. If so, save these flows in $G'$ and subtract them in the current group $G$. Likewise, in lines~22-23, subtract the completed direct flows and relay flows in the unscheduled set $V_d$ and $V_r$, respectively. Then the algorithm will find the flows that are nonadjacent to ongoing flows in $G$ by calling functions $DGroup$ and $RGroup$, to a new group $G$, in which flows can transmit concurrently. We save this group in $\mathbb{G}$. At the end of each slot, as shown in line~27, we check if all the flows have been accomplished. If so, break the iteration and $m$ is the actual time slot consumption; otherwise, the loop continues. After finishing the flow transmission, we calculate the system throughput in line 29. The computational complexity of Algorithm~\ref{alg:JRDS} is on the order of ${\color{black}\mathcal{O} (N^2)}$.

\begin{figure}[!t]
\vspace*{-1mm}
\begin{center}
\includegraphics[width=0.95\columnwidth]{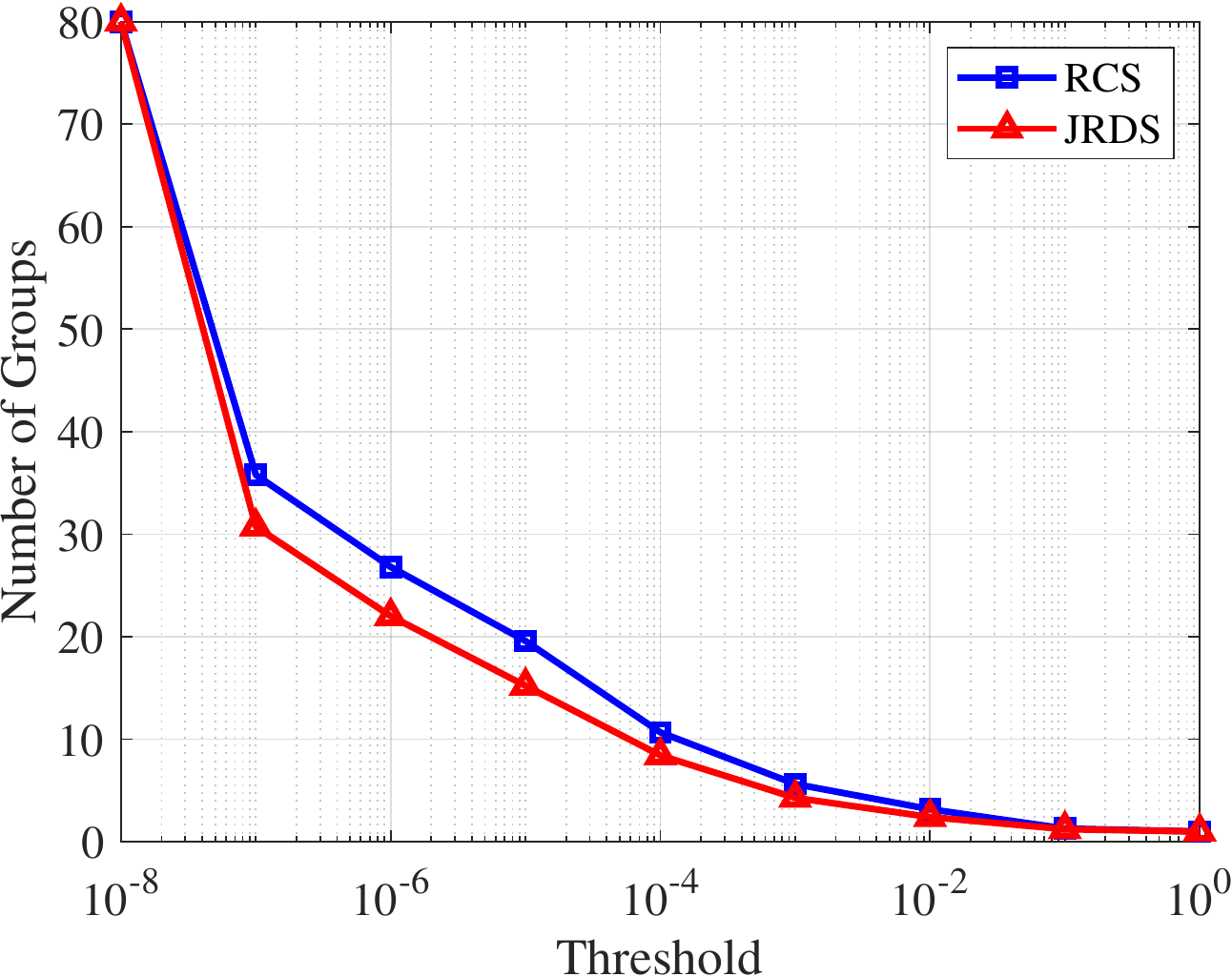}
\end{center}
\vspace*{-4mm}
\caption{Comparison of numbers of groups as the functions of threshold $\sigma$ for RCS and JRDS under the default system setting of Subsection~\ref{S5.1}.}
\label{fig:Number of groups} 
\vspace*{-4mm}
\end{figure}

We compare the numbers of groups obtained by RCS and JRDS in Fig.~\ref{fig:Number of groups} under the default system setting of Subsection~\ref{S5.1}. As can be seen from Fig.~\ref{fig:Number of groups}, the number of groups obtained by JRDS is smaller than that of RCS, especially for $\sigma$ in the range of $10^{-7}$ to $10^{-4}$. This is because the joint scheme avoids collisions effectively and hence more flows can be divided into the same group. 

\section{Performance Evaluation}\label{S5}

\subsection{Simulation System Setup}\label{S5.1}

{\color{black}Consider a straight 3-lane highway scenario of 6\,km length served by an RSU and 5 UAVs, as illustrated in Fig.~\ref{fig:Scenario}.  
The number of vehicles at each lane is set to 60, with the average speed of $100$\,km/h. Note that the safety distance between adjacent vehicles is satisfied.}
We assume that the transceivers for each flow are selected randomly, and the throughput requirement of each flow ranges in $[0.1, ~  1]$\,Gbps. In the simulation, the positions of vehicles and UAVs are updated by slot, and dynamic scheduling algorithms are executed. Unless otherwise stated, the default simulation parameters listed in Table~\ref{table:Initial_Parameters} are used, also see \cite{WYB,Antenna,3GPP2}.

\begin{table}[!t] 
\vspace*{-1mm}
\caption{Default System Parameters}
\label{table:Initial_Parameters} 
\vspace*{-4mm}
\begin{center}
\begin{tabular}{lcl}  
\toprule   
  Parameter              & Symbol              & Value \\ \midrule 
  {\color{black}Carrier frequency}      & {\color{black}$f$}      & {\color{black}30 GHz}\\
  Number of flows        & ${\color{black}N}$                 & 80 \\ 
  Number of time slots   & $M$                 & 2000 \\  
  Slot duration          & $T$                 & 0.1 s \\
  Fading depth           & $m$                 & 2 \\
  Background noise       & $N_0$               & -134 dBm/MHz \\    
  System bandwidth       & $W$                 & 2000 MHz \\
  Transmission power     & $P_t$               & 40 dBm\\
  Average power of UAV   & $\overline{P_{u}}$  & 30 dBm\\
  Peak power of UAV      & $\widetilde{P_{u}}$ & 2$\overline{P_{u}}$ \\
  Transceiver efficiency & $\eta$              & 0.8 \\  
  Height of UAV          & $h_u$                 & 100 m \\
  Moving speed of UAV    & $V_u$               & 20 m/s \\
  PL factor for V2V      & $\alpha_v$          & 2.5 \\
  PL factor for U2V      & $\alpha_u$          & 2 \\
  Rician K-factor     & $\mathcal{K}$       &  9 dB \\
  Interference threshold & $\sigma$            & $10^{-3}$ \\
  SI cancellation level  & $\beta$             & $10^{-9}$ \\
  Maximum antenna gain   & $G_{\color{black} 0}$ & 20 dBi \\
  Half-power beamwidth   & $\theta_{\rm 3dB}$  & $30^\circ$ \\
\bottomrule  
\end{tabular}
\end{center}
\vspace*{-5mm}
\end{table}

Besides, two mobility models are considered in the simulation, the Poisson process (PP) and interrupted PP (IPP). The PP model has been introduced in Section~\ref{S2}. For the IPP, the inter-arrival distance between two consecutive vehicles follows a second-order hyper-exponential distribution with the mean of $E(d_n)\! =\! \frac{p}{\overline{v}_1\cdot 2\,\text[s]}+\frac{(1-p)}{\overline{v}_2\cdot 2\,\text[s]}$ \cite{IPP}. Unless otherwise specified, the mobility model is PP. 

{\color{black}
To show the advantages of our proposed RCS and JRDS algorithms in improving performance, we compare them with three other schemes, namely, the exhaustive method which obtains the optimal solution, the random relay concurrent transmission (RR) scheme that randomly selects relays \cite{Ding}, and the serial TDMA scheme. Two performance metrics considered are:}
\begin{enumerate}
\item \textbf{Transmission Time}: The total number of time slots consumed to complete flow transmission.
\item \textbf{Network Throughput}: The achieved throughput of completed flows in the network [Gbps]. 
\end{enumerate}

\begin{figure}[!ht]
\subfloat[][]{
\includegraphics[width=0.95\columnwidth]{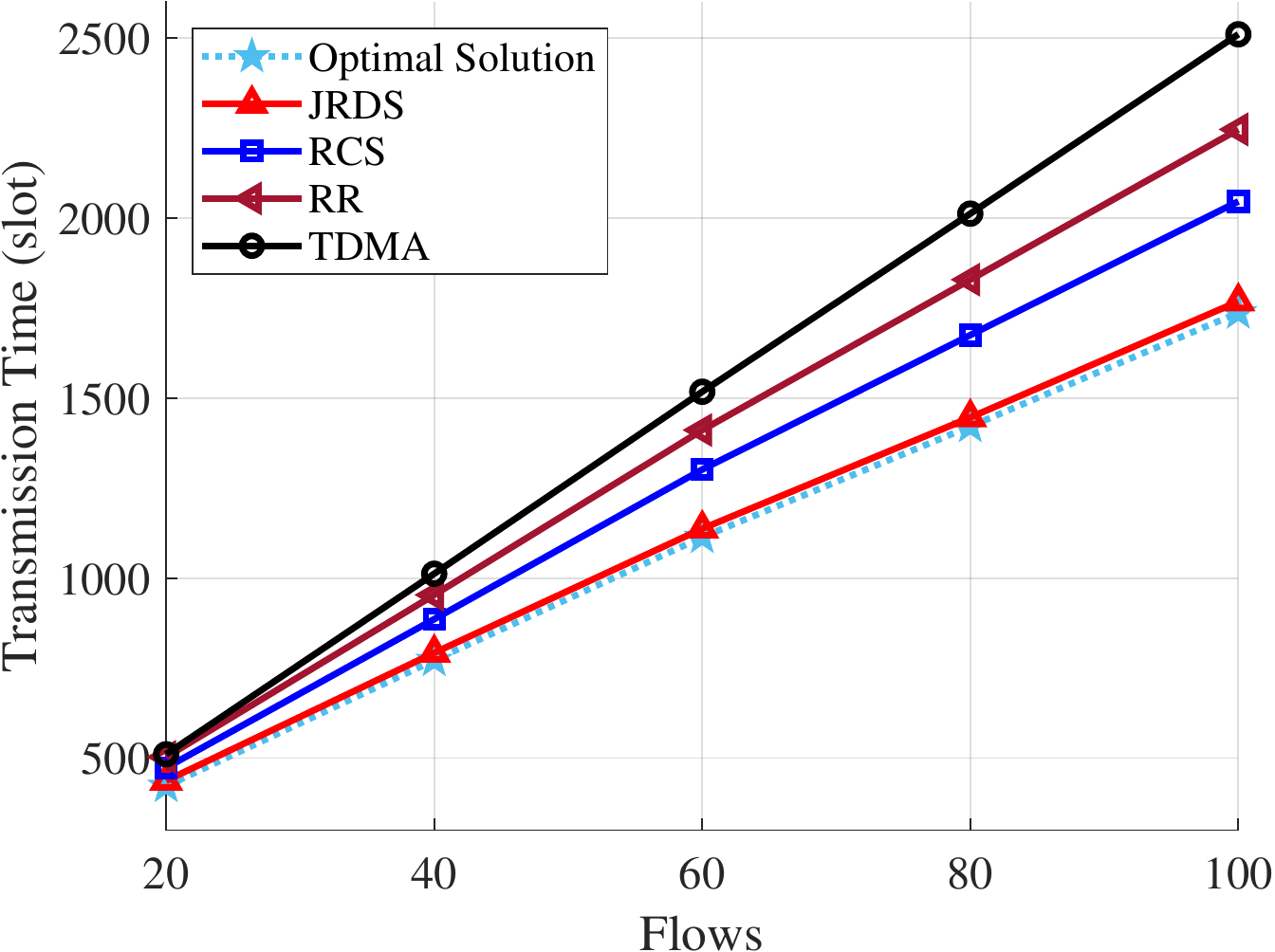}
 \label{fig:8a}} 
\vspace*{-2mm} \\
\subfloat[][]{
\includegraphics[width=0.95\columnwidth]{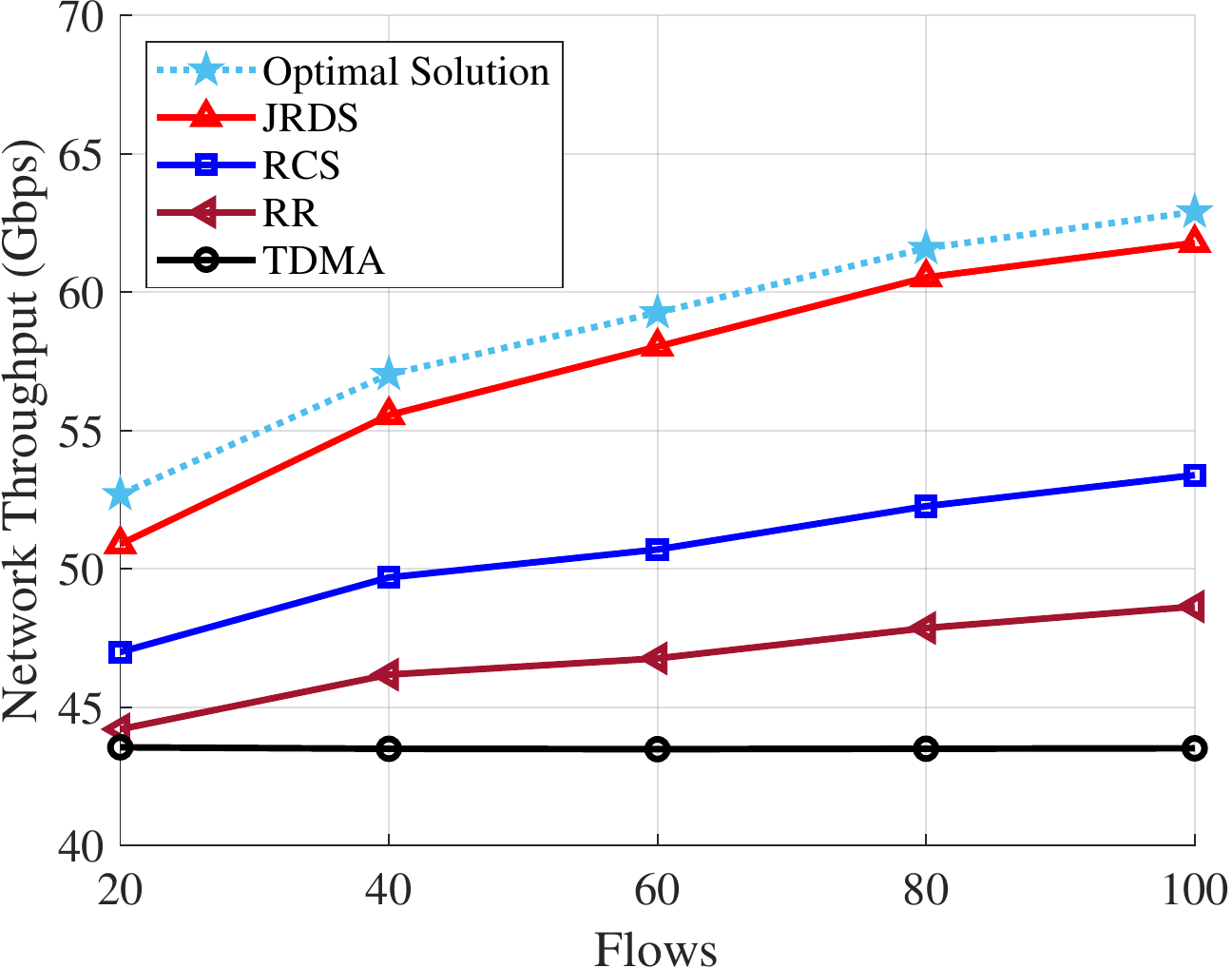}
\label{fig:8b}} 
\vspace*{-1mm}
\caption{Performance comparison of the optimal solution with four schemes: (a)~transmission time, and (b)~network throughput.}
\label{fig:Compare} 
\vspace*{-4mm}
\end{figure}
	
\subsection{Comparison with Optimal Solution}\label{S5.2}

Fig.~\ref{fig:Compare} compares the performance of our RCS and JRDS as well as the benchmarks TDMA and RR with the optimal solution of the optimization problem P1, given different numbers of flows. Since obtaining the optimal solution is extremely time-consuming, the simulation is performed only for up to 100 flows.
As can be observed, JRDS offers near-optimal performance, which outperforms RCS, RR and TDMA schemes. 
In terms of transmission time, the gap between the optimal solution and our JRDS scheme stays negligible under different numbers of flows, as shown in Fig.~\ref{fig:Compare}\,(a). 
As can be seen from Fig.~\ref{fig:Compare}\,(b), the gap in network throughput between the optimal solution and our JRDS is small and decreases with the increase of flow number. When $N=100$, the gap is only around {\color{black}2.4\%}. 
{\color{black} Since the performance of JRDS is so close to that of the optimal solution, it is sufficient to compare JRDS with RCS and other existing schemes in the following simulations.}

\begin{figure}[!t]
\vspace*{-2mm}
\subfloat[][]{
\includegraphics[width=0.95\columnwidth]{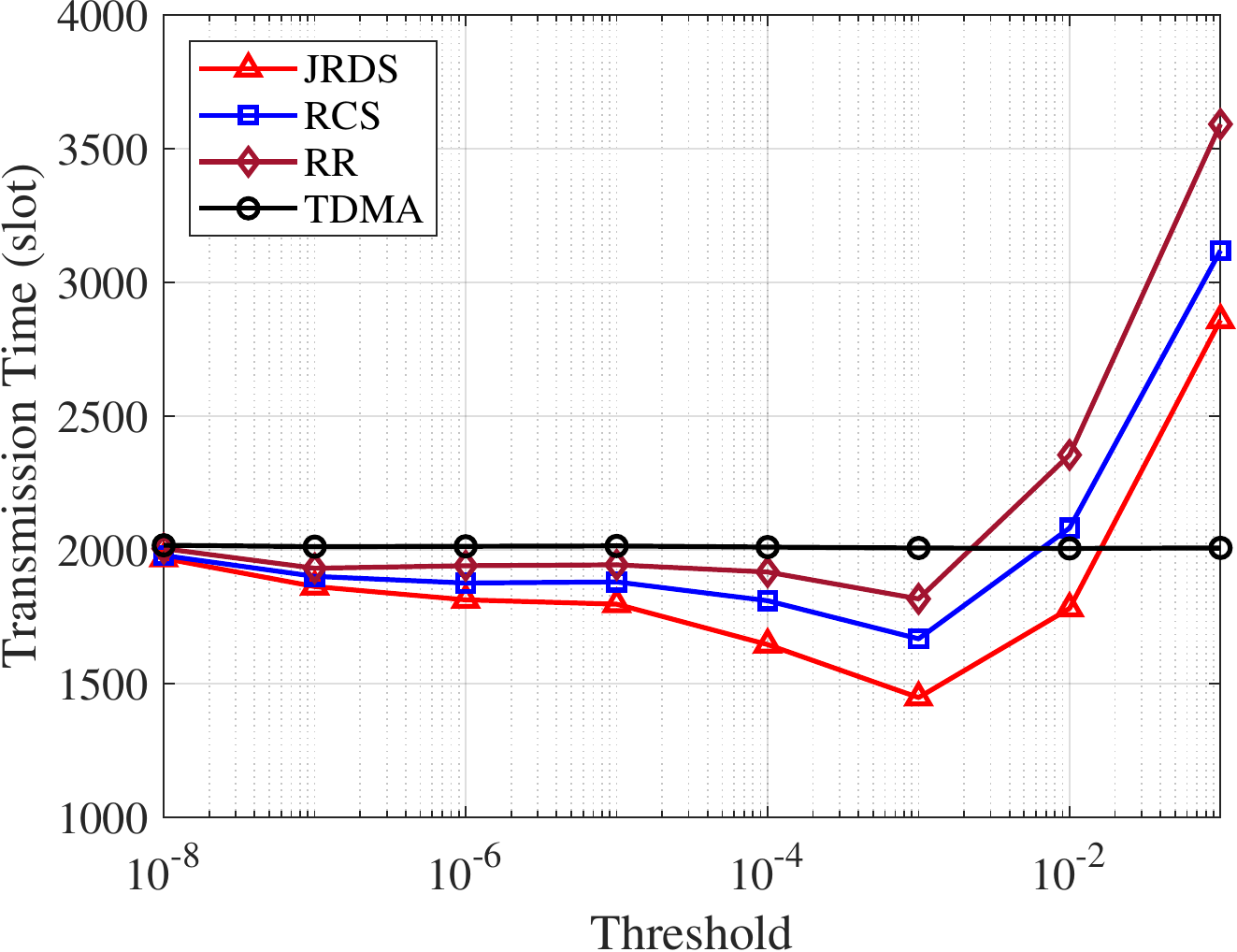}
 \label{fig:9a}}  
\vspace*{-2mm} \\
\subfloat[][]{
\includegraphics[width=0.95\columnwidth]{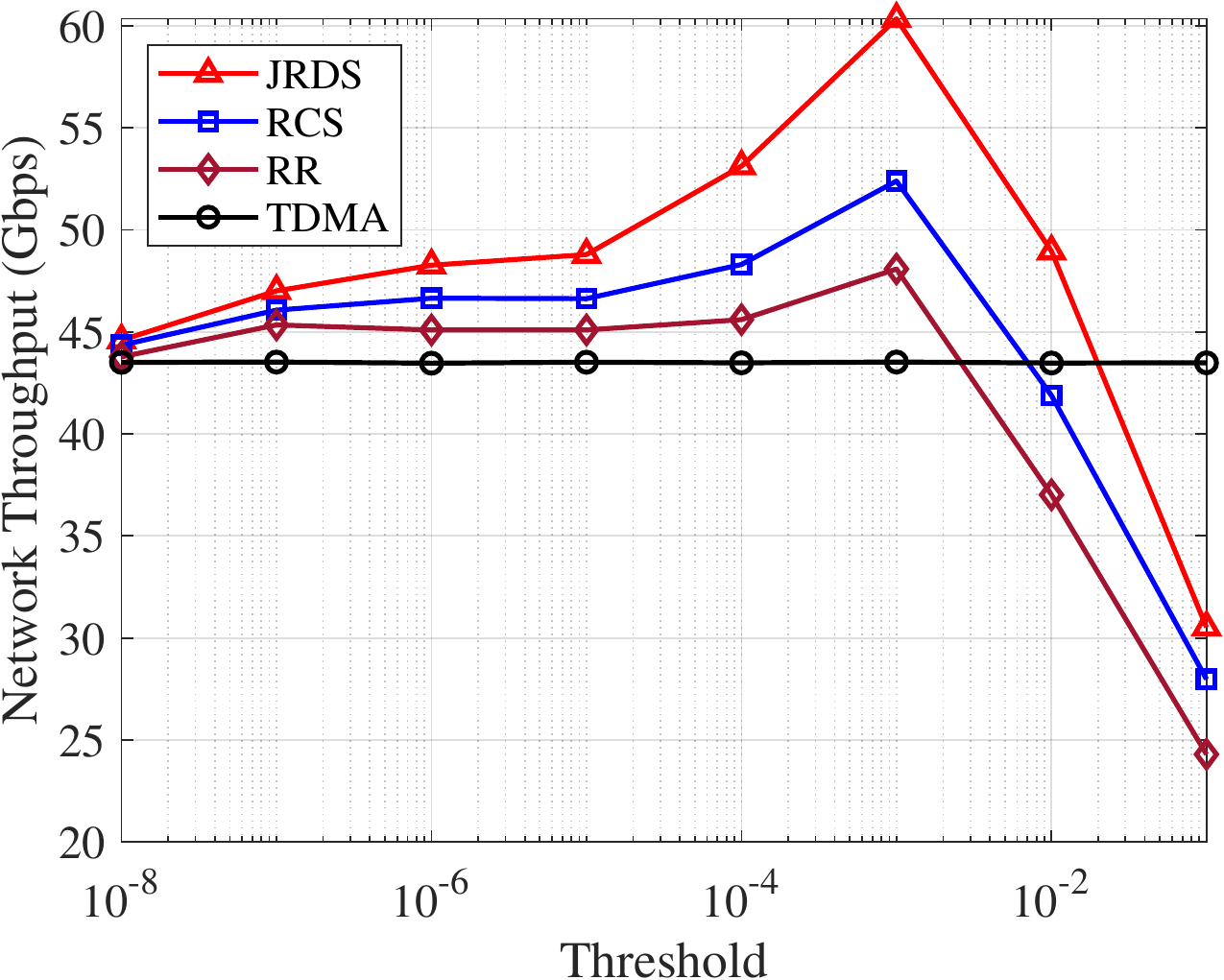}
\label{fig:9b}}  
\vspace*{-1mm}
\caption{Influence of threshold $\sigma$ to the achievable performance of four schemes: (a)~transmission time, and (b)~network throughput.}
\label{fig:Threshold}  
\vspace*{-4mm}
\end{figure}

\subsection{Choice of Interference Threshold}\label{S5.3}

The influence of the interference threshold $\sigma$ on the achievable performance of the {\color{black}four} schemes is investigated in Fig.~\ref{fig:Threshold}. {\color{black}Obviously, the choice of $\sigma$ does not affect the performance of TDMA, since it leverages serial transmission without flow contention.} But for {\color{black} RR}, RCS and JRDS, as $\sigma$ changes from $10^{-8}$ to $10^{-3}$, the achievable performance improves gradually. This is because increasing the threshold enable more flows to transmit concurrently, therefore reducing transmission time and enhancing overall throughput. {\color{black} After reaching the best performance at $\sigma\! =\! 10^{-3}$, further increasing $\sigma$ leads to the performance degradation. This is due to the fact that with a too large threshold, severe interference exists between concurrent flows, which can reduce the data rate of each flow to be smaller than that under the serial TDMA scheme. Thus, unless otherwise specified, we fix $\sigma\! =\! 10^{-3}$ in the evaluation.}

{\color{black} From the results of Fig.~\ref{fig:Threshold}, we can also see that our two schemes consume less transmission time and yet achieve higher throughput compared with RR, thanks to the proposed relay selection scheme, which 1) selects relays of high probability to enable LoS transmission, and therefore contribute to reducing the time consumption; and 2) avoids different blocked flows to select the same relay, and hence is beneficial to exploit concurrent transmissions for enhancing the performance. In addition, JRDS outperforms RCS considerably because dynamic scheduling is leveraged.}

\begin{figure}[!ht]
\vspace*{1mm}
\begin{center}
\includegraphics[width=0.95\columnwidth]{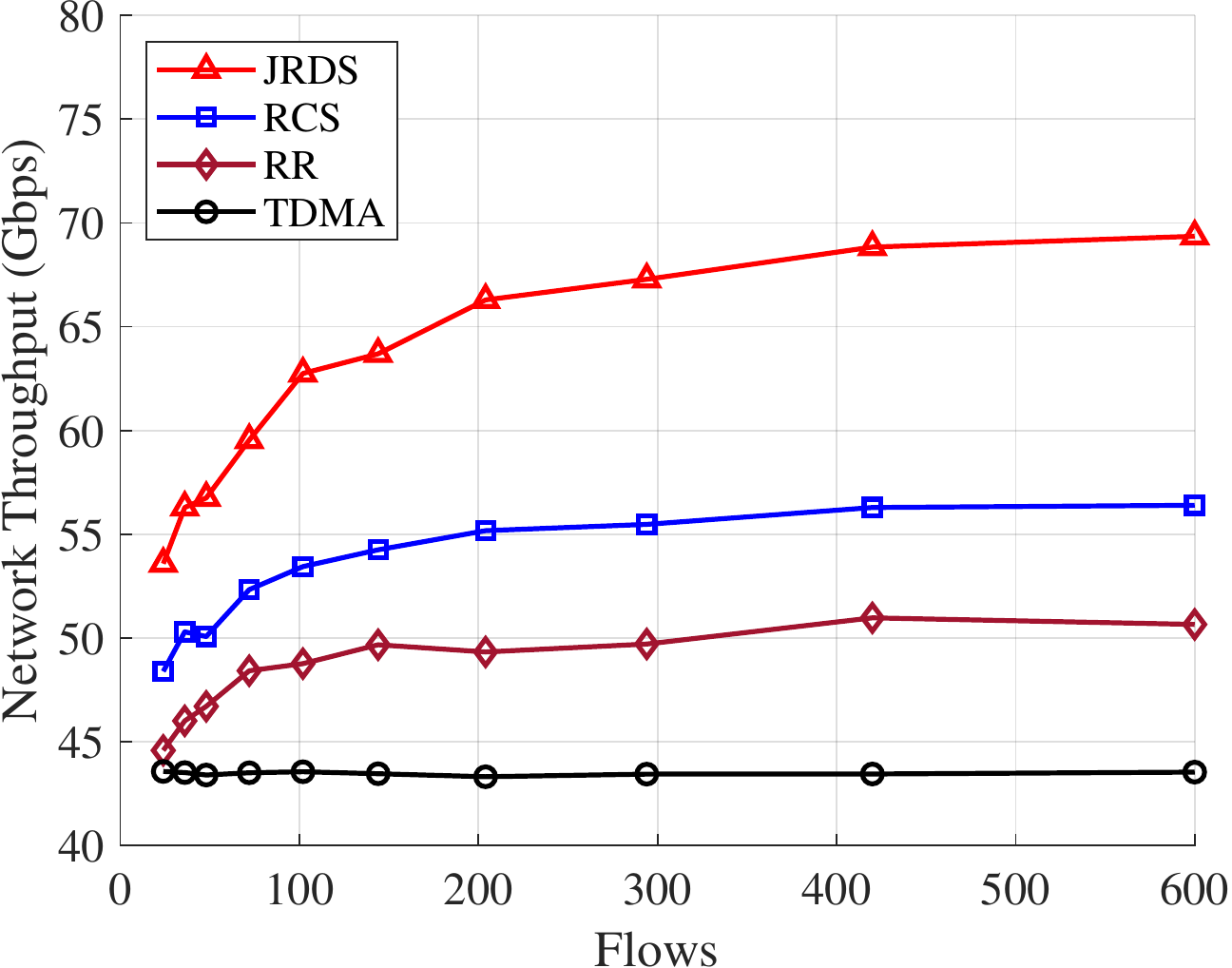}
\end{center}
\vspace*{-4mm}
\caption{Network throughput comparison of four schemes given different numbers of flows.}
\label{fig:ten} 
\vspace*{-4mm}
\end{figure}	
\subsection{Performance under Different Numbers of Flows}\label{5.4}
  
In Fig.~\ref{fig:ten}, we plot the network throughput of these four schemes given different numbers of flows. Observe that the curve of TDMA stays flat. This is because TDMA adopts serial transmission, and the number of completed bits or flows is proportional to the transmission time. Hence its throughput is not depending on ${\color{black}N}$. By contrast, as ${\color{black}N}$ increases, the network throughput of RR, RCS and JRDS first increase as well until reaching the saturation values at {\color{black}$N\! =\! 150$ for RR,} ${\color{black}N}\! =\! 300$ for RCS and ${\color{black}N}\! =\! 400$ for JRDS, respectively. Evidently, concurrent transmission enables {\color{black}RR,} RCS and JRDS to achieve much higher network throughput than TDMA. Also as expected, JRDS attains the best performance and it significantly outperforms the second best RCS.

\begin{figure}[!ht]
\subfloat[][]{
  \includegraphics[width=0.95\columnwidth]{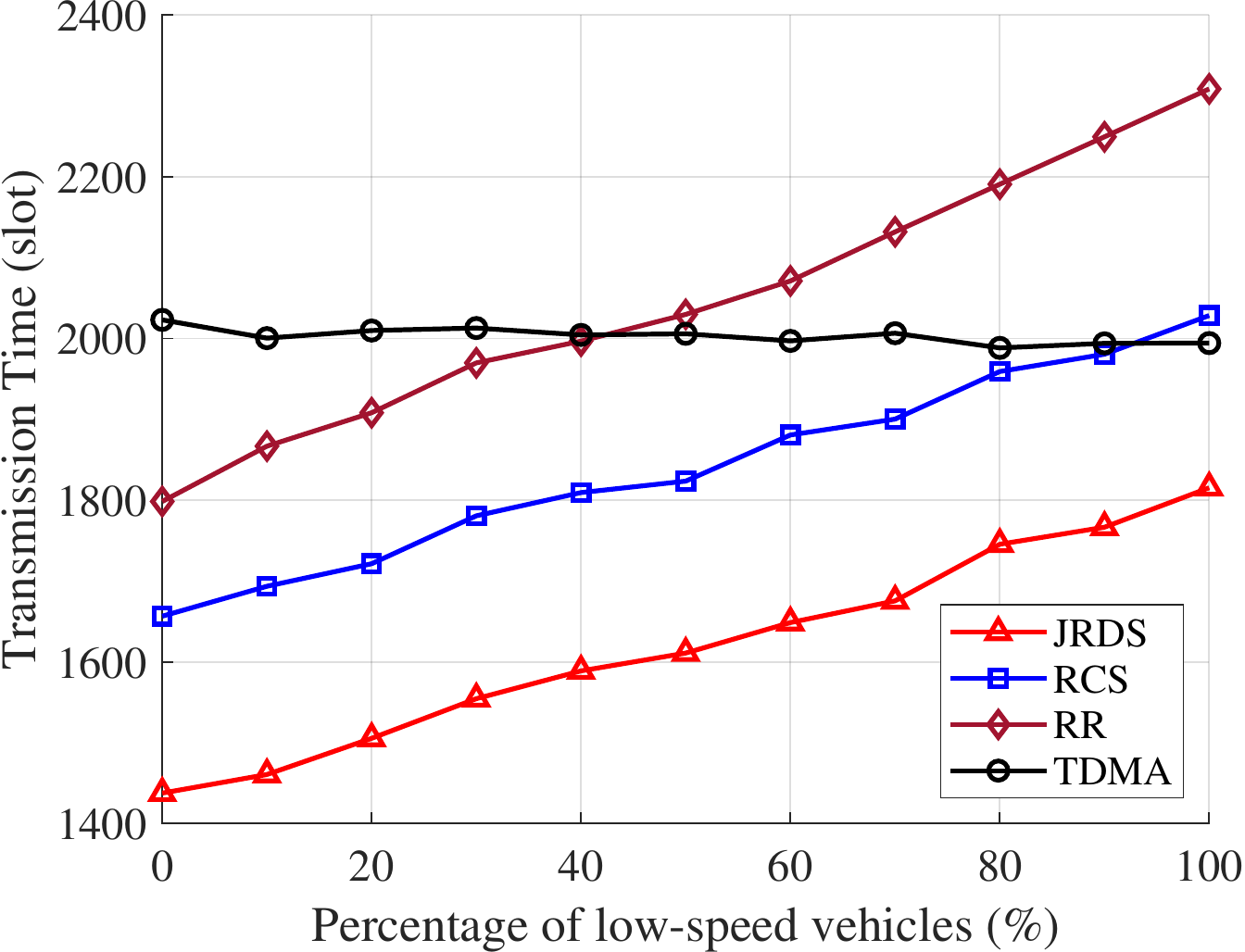}\label{fig:elea}}
  \vspace*{-2mm} \\
\subfloat[][]{
  \includegraphics[width=0.95\columnwidth]{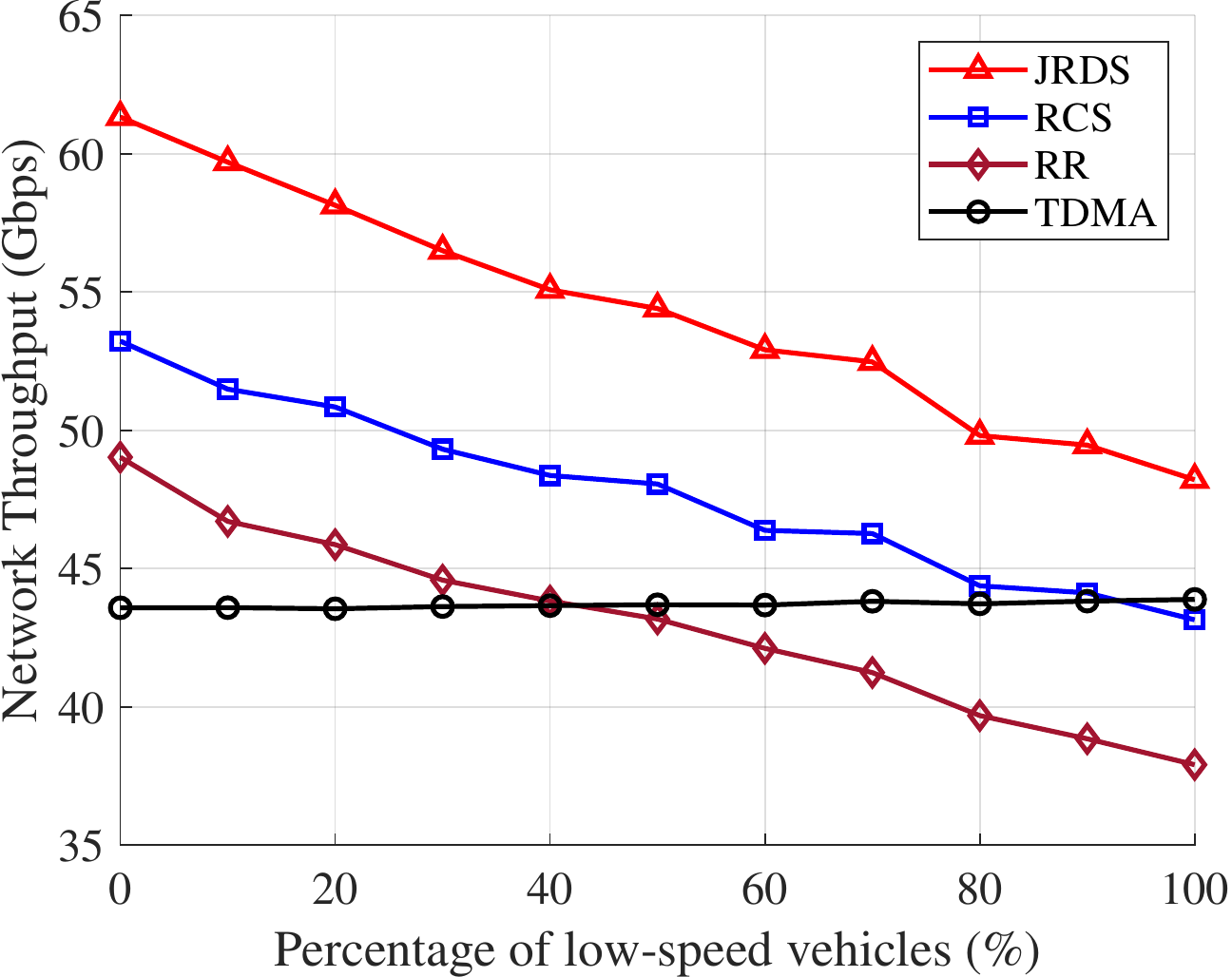}\label{fig:eleb}}
  \vspace*{-1mm}
\caption{{\color{black}Influence of mobility model to the achievable performance of four schemes: (a)~transmission time, and (b)~network throughput.}}
\label{fig:ele} 
\vspace*{-4mm}
\end{figure}

\subsection{Performance under IPP Traffic Model}\label{S5.5}

We also investigate the impact of mobility model on system performance. Specifically, in this experiment, we adopt
an IPP model with $\overline{v}_1=60$\,km/h and $\overline{v}_2=100$\,km/h. Fig.~\ref{fig:ele} compares the performance of the four schemes in terms of time slot consumption and network throughput under increasing percentage of low-speed vehicles.

It can be seen from Fig.~\ref{fig:ele}\,(a) that JRDS consumes the lowest transmission time, which is expected. As the percentage of low-speed vehicles increases, the transmission distance of flows decreases but the interference among them increases. Because the negative impact of the latter outweighs the beneficial effect of the former, the time slot consumption of RR, RCS and JRDS increases linearly with the growth of low-speed vehicles. By contrast, the serial TDMA is less affected by the interference. Therefore, when the transmission distance of flows decreases, the received power increases. This leads to a slightly improvement in its performance.
Similarly, Fig.~\ref{fig:ele}\,(b) confirms that JRDS attains the highest network throughput. Also the network throughput of RR, RCS and JRDS decreases linearly with increasing percentage of low-speed vehicles but that of TDMA shows a slightly improvement. 

Fig.~\ref{fig:ele} also shows that when the percentage of low-speed vehicles exceeds 40\% and 90\%, TDMA outperforms RR and RCS, respectively. In this case, the enforced flow grouping increases transmission interference in groups and degrades the performance of concurrent transmission. However, our JRDS always outperforms TDMA significantly even when the percentage of low-speed vehicles reaches 100\%.

\begin{figure}[!htp]
\vspace*{-6mm}
\subfloat[][]{
  \includegraphics[width=0.95\columnwidth]{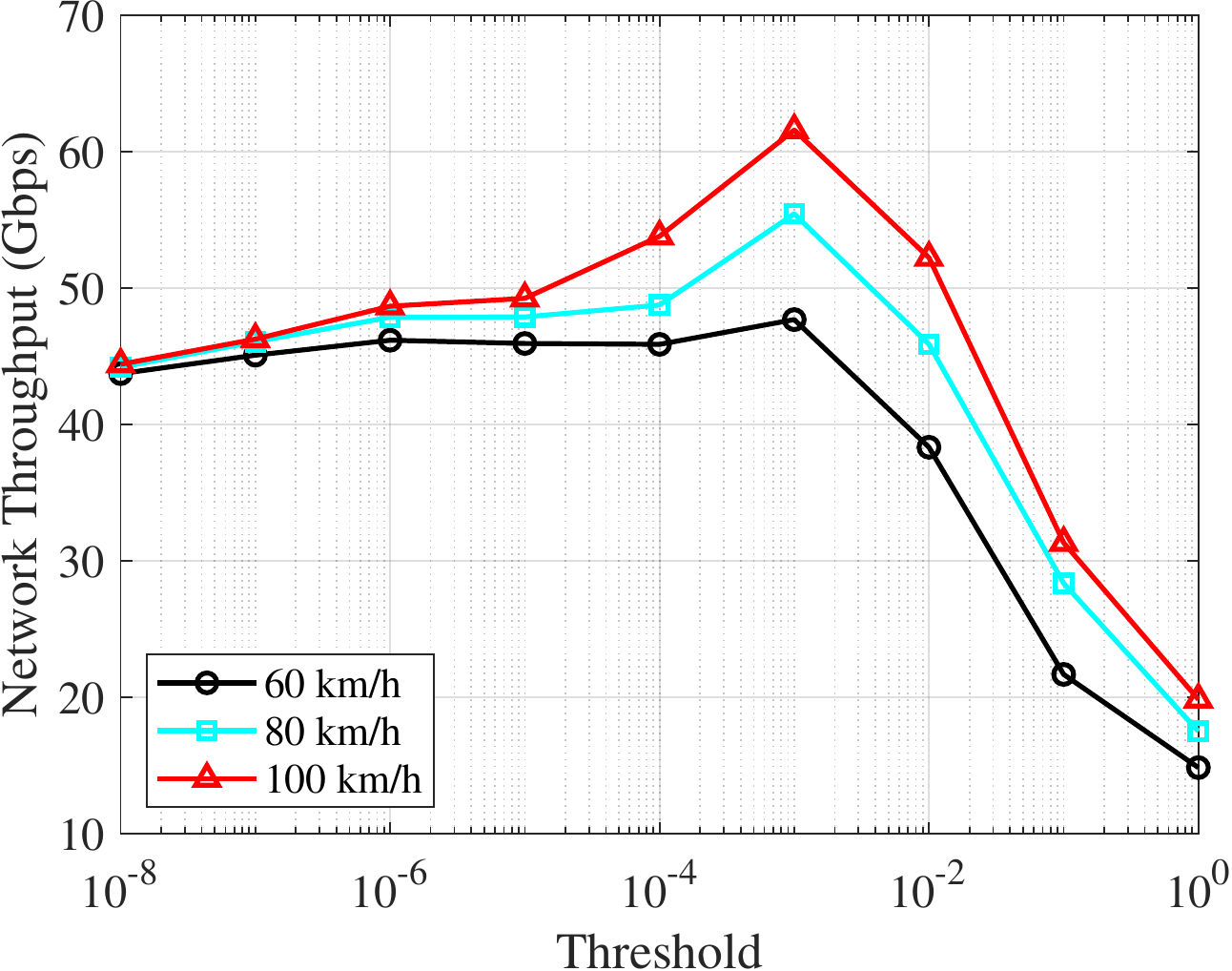}\label{fig:twea}}
  \vspace*{-1mm} \\
  \subfloat[][]{
  \includegraphics[width=0.95\columnwidth]{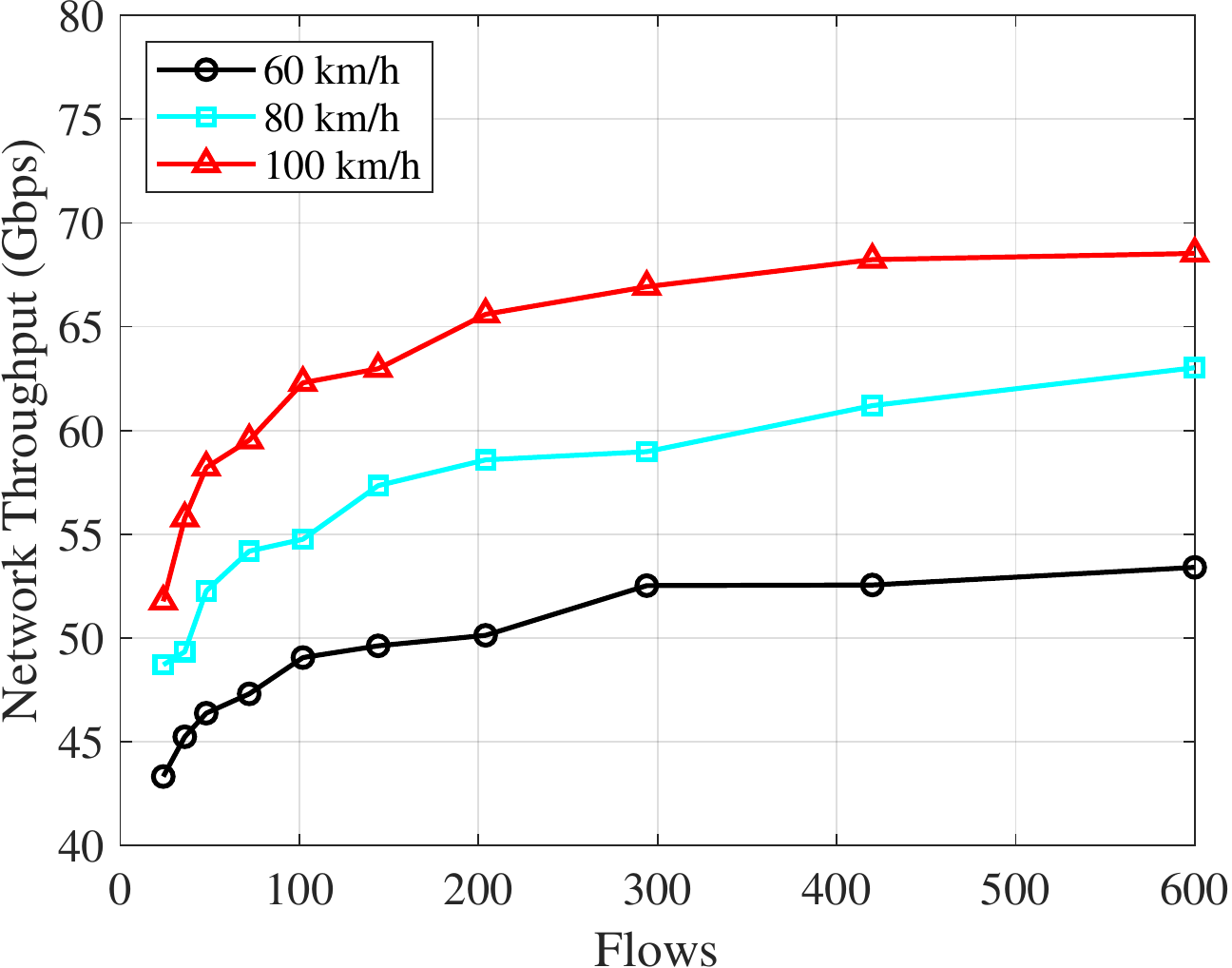}\label{fig:tweb}}
	\vspace*{-1mm}
\caption{{\color{black}Influence of mobility speed $\overline{v}$ to the network throughput of JRDS: (a)~under different thresholds, and (b)~under different numbers of flows.}}
\label{fig:twe} 
\vspace*{-4mm}
\end{figure}
\subsection{Performance of JRDS under Different Speeds}\label{S5.6}

As demonstrated in the previous results, JRDS outperforms the other schemes in the dynamic scenario. Here, we focus on the performance of JRDS under different speeds. Three cases of $\overline{v}$, 60\,km/h, 80\,km/h, and 100\,km/h, are investigated.

Fig.~\ref{fig:twe}\,(a) plots the network throughput of JRDS as the function of the interference threshold $\sigma$ given three different speeds. It is noted that JRDS achieves better performance at higher speed. The reason is that at higher speed, the interference power decreases significantly, which offsets the power loss of V2V links due to the increased transmission distance. Also observe that when $\sigma$ ranges from $10^{-8}$ to $10^{-5}$, the throughput at different speeds is close since concurrent transmission is not fully leveraged. The optimal threshold is $\sigma\! =\! 10^{-3}$, which is consistent with the results in Fig.~\ref{fig:Threshold}.

In Fig.~\ref{fig:twe}\,(b), we plot the achievable network throughput of JRDS as the function of the number of flows $N$ given three different speeds. Again we observe that the dynamic scheduling of JRDS achieves better performance at higher speed, which implies that V2V links experience severe interference when users are distributed densely. 

\section{Conclusions}\label{S6}

In this paper, we have formulated the joint optimization of relay selection and transmission scheduling in a UAV-aided mmWave vehicular network, to minimize transmission time while meeting the throughput requirements. Since the optimization problem is NP-hard, we have designed two heuristic schemes, called RCS and JRDS, to leverage  the advantages of concurrent transmission. RCS randomly selects relay in the candidate relay set for each blocked flow and enables concurrent transmission. It is of low-complexity and achieves better performance than the existing RR and TDMA schemes. JRDS executes dynamic relay selection with transmission scheduling, which considerably reduces both relay collisions and the interference level in concurrent transmissions. The simulation results have demonstrated that JRDS offers near optimal solution to the NP-hard joint optimization of relay selection and transmission scheduling. Extensive simulations have also been conducted to investigate the impact of various system parameters, including the choice of interference threshold, arrival speed, and network load, in order to provide useful guidelines for the system design. {\color{black} Our future work will study how to minimize the scheduling delay while maintaining the beneficial advantages of JRDS in reducing the transmission time and enhancing network throughput.}
 
\bibliographystyle{IEEEtran}

\end{document}